\documentclass[sigconf]{acmart}

\usepackage{stfloats} 
\usepackage{balance}
\AtBeginDocument{%
  }

\renewcommand\footnotetextcopyrightpermission[1]{} 
\copyrightyear{2023} 
\acmYear{2023} 
\setcopyright{acmlicensed}\acmConference[MM '23]{Proceedings of the 31st ACM International Conference on Multimedia}{October 29-November 3, 2023}{Ottawa, ON, Canada}
\acmBooktitle{Proceedings of the 31st ACM International Conference on Multimedia (MM '23), October 29-November 3, 2023, Ottawa, ON, Canada}
\acmPrice{15.00}
\acmDOI{10.1145/3581783.3612209}
\acmISBN{979-8-4007-0108-5/23/10}

\thanks{\copyright {Owner/Author | ACM} {2023}. This is the author's version of the work. We have updated the experimental results that were necessary due to the updated dataset. This has been done to avoid any disclosure of sensitive information. It is posted here for your personal use. Not for redistribution. The definitive Version of Record was published in {Proceedings of the 31st ACM International Conference on Multimedia (MM '23), October 29-November 3, 2023, Ottawa, ON, Canada}, http://dx.doi.org/10.1145/3581783.3612209.}

\begin{document}

\title{MORE: A Multimodal Object-Entity Relation Extraction Dataset with a Benchmark Evaluation}

\author{Liang He}
\affiliation{%
  \institution{National Key Laboratory for Novel Software Technology}
  \institution{Nanjing University}
  \city{Nanjing}
  \country{China}
}
\email{heliang@smail.nju.edu.cn}

\author{Hongke Wang}
\affiliation{%
  \institution{National Key Laboratory for Novel Software Technology}
  \institution{Nanjing University}
  \city{Nanjing}
  \country{China}
}
\email{wanghk@smail.nju.edu.cn}

\author{Yongchang Cao}
\affiliation{%
  \institution{National Key Laboratory for Novel Software Technology}
  \institution{Nanjing University}
  \city{Nanjing}
  \country{China}
}
\email{caoyc@smail.nju.edu.cn}

\author{Zhen Wu}
\authornote{Corresponding author.}
\affiliation{%
  \institution{National Key Laboratory for Novel Software Technology}
  \institution{Nanjing University}
  \city{Nanjing}
  \country{China}
}
\email{wuz@nju.edu.cn}

\author{Jianbing Zhang}
\affiliation{%
  \institution{National Key Laboratory for Novel Software Technology}
  \institution{Nanjing University}
  \city{Nanjing}
  \country{China}
}
\email{zjb@nju.edu.cn}

\author{Xinyu Dai}
\affiliation{%
  \institution{National Key Laboratory for Novel Software Technology}
  \institution{Nanjing University}
  \city{Nanjing}
  \country{China}
}
\email{daixinyu@nju.edu.cn}

\renewcommand{\shortauthors}{Liang He et al.}



\begin{abstract}
Extracting relational facts from multimodal data is a crucial task in the field of multimedia and knowledge graphs that feeds into widespread real-world applications. 
The emphasis of recent studies centers on recognizing relational facts in which both entities are present in one modality and supplementary information is used from other modalities. However, such works disregard a substantial amount of multimodal relational facts that arise across different modalities, such as one entity seen in a text and another in an image. 
In this paper, we propose a new task, namely Multimodal Object-Entity Relation Extraction, which aims to extract "object-entity" relational facts from image and text data. To facilitate research on this task, we introduce MORE, a new dataset comprising 21 relation types and 20,264 multimodal relational facts annotated on 3,559 pairs of textual news titles and corresponding images. To show the challenges of Multimodal Object-Entity Relation Extraction, we evaluated recent state-of-the-art methods for multimodal relation extraction and conducted a comprehensive experimentation analysis on MORE. Our results demonstrate significant challenges for existing methods, underlining the need for further research on this task. Based on our experiments, we identify several promising directions for future research. The MORE dataset and code are available at \url{https://github.com/NJUNLP/MORE}.

\end{abstract}




\maketitle

\section{Introduction}

Relation extraction (RE) plays a critical role in constructing knowledge graphs, and with the abundance of multimedia data resources, such as web images, many researchers have explored the ability to extract entities and predict their relations from multimodal data \cite{DBLP:conf/cvpr/TangNHSZ20, DBLP:conf/icmcs/ZhengWFF021, DBLP:conf/aaai/WanZDHYP21}. This approach has become increasingly relevant in building or populating knowledge graphs, especially multimodal knowledge graphs \cite{DBLP:conf/esws/LiuLGNOR19, DBLP:conf/semweb/FerradaBH17, DBLP:journals/bdr/WangWQZ20}, which have proven useful in downstream tasks, including question answering \cite{DBLP:conf/semweb/FerradaBH17a, DBLP:conf/naacl/MitraRS22}, recommendation \cite{DBLP:conf/i-semantics/LullyLSR18, wu2022state}, and reasoning \cite{liang2022reasoning, DBLP:conf/icde/Zheng0QYCZ23}. 
Recently, Zheng et al. \cite{DBLP:conf/icmcs/ZhengWFF021} introduces the concept of multimodal relation extraction (MRE), which identifies textual relations using visual clues. Such works provide valuable insights and achieve promising results in this field.

\begin{figure}[ht]
    \centering
    \includegraphics[width=0.75\linewidth]{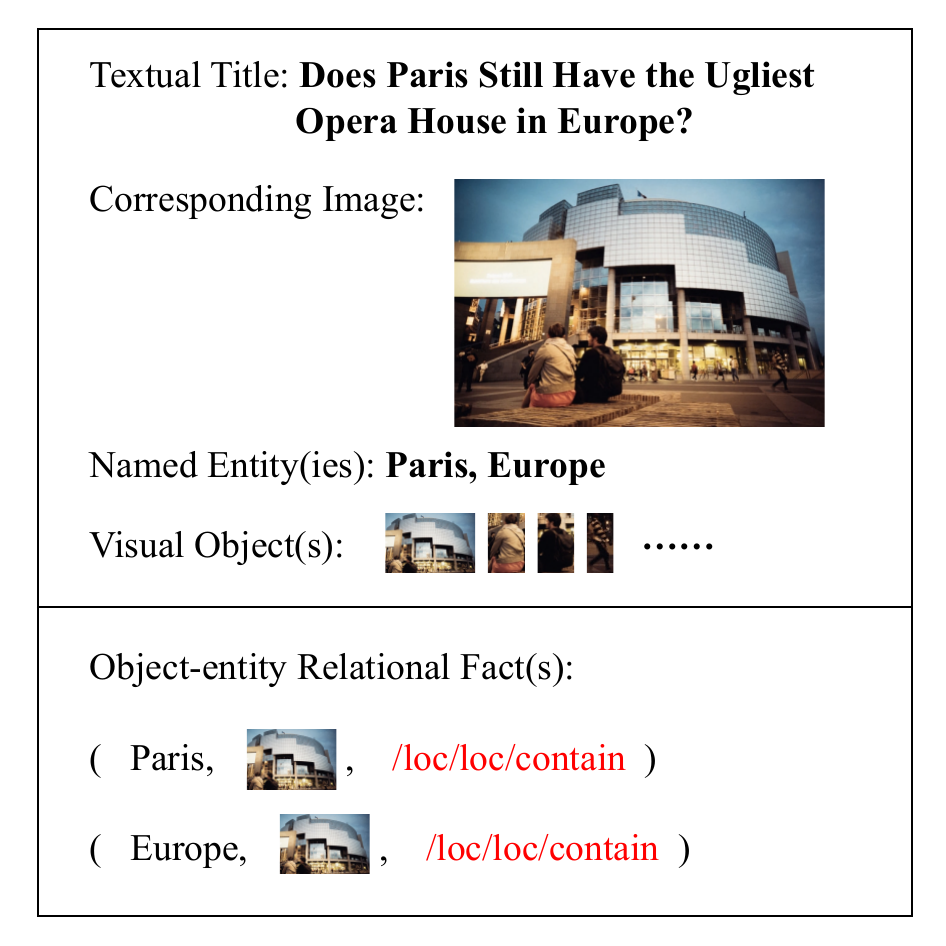}
    \caption{An example from MORE, including two entities in the text, and several visual objects from the image, yielding two new multimodal relational facts. 
    }
    \label{fig:example}
\end{figure}

Despite the successful efforts in relation extraction from multimodal data, there is an inevitable restriction in practice. 
Recent works focus primarily on identifying relational facts for cases where both entities appear in the same modality, with other modalities serving mainly as sources of supplementary information. However, in practical applications, a significant number of multimodal relational facts come from data across multiple modalities e.g., one entity may appear in the text and another in the image. A key challenge in these cases is to accurately identify the visual object to which a given text refers in order to correctly establish a new relational fact for the "object-entity" pair.
Figure \ref{fig:example} shows an example of this kind of multimodal relational facts. It is difficult to identify the relational facts of such object-entity pairs with conventional MRE tasks. The statistics on our human-annotated corpus sampled from 
online news indicate that object-entity pairs originating from separate modalities account for at least 30\% of the total, which is not negligible.

In this paper, we introduce a new task called \textbf{Multimodal Object-Entity Relation Extraction}, which aims to extract object-entity relational facts from text and image data. In real-world data, such as news titles, the text does not necessarily directly describe accompanying images. This challenges the model to account for the inconsistencies in semantics between textual and visual data. Additionally, images can contain multiple objects, making it challenging to identify which object is being referred to in the text. Lastly, there is an urgent need for a large-scale, manually-annotated, and general-purpose dataset for this new task.

Therefore, we present \textbf{MORE}, a new Multimodal Object-Entity Relation Extraction dataset, which has the following three distinct features. First, MORE contains 21 relation types, 13,520 visual objects, and 20,264 multimodal relational facts annotated on 3,559 pairs of textual news titles and corresponding images, rendering it unique and larger than previous MRE datasets. Second, all the facts involve the extraction of entities from texts and objects from images, requiring a deep understanding of their interactions. Third, there are an average of 3.8 objects per image in MORE, presenting a higher challenge to models trying to extract the correct object.

In addition to constructing the dataset, to assess the challenges of Multimodal Object-Entity Relation Extraction, we implemented state-of-the-art MRE methods and adapted them for this new task. We conducted a comprehensive evaluation of all these models using our dataset, and the results reveal that the performance of the existing methods declined significantly on MORE. This finding highlights that Multimodal Object-Entity Relation Extraction is more challenging and remains an open problem.

Additionally, we propose \textbf{MOREformer} to improve visual-textual understanding, including attribute-aware textual encoding, depth-aware visual encoding, and position-fused multimodal encoding. A detailed analysis of the results also shows the effectiveness of this method, indicating that incorporating visual-textual interaction is promising and worthy of further research.

In summary, our contribution can be summarized as three-fold: 

\begin{itemize}

\item Firstly, we introduce a new task named Multimodal Object-Entity Relation Extraction and correspondingly propose a new dataset called MORE. This dataset contains a large number of object-entity relational facts and presents new challenges to the extraction of multimodal facts from both text and image data. 

\item Next, we systematically adapt the most recent state-of-the-art MRE methods for Multimodal Object-Entity Relation Extraction. In addition, we propose MOREformer to explore ways of encoding textual attributes, visual depth, and positions that could potentially be valuable for future research.

\item Finally, we conduct a comprehensive evaluation and in-depth analysis of the proposed methods on our dataset, indicating promising research directions for Multimodal Object-Entity Relation Extraction.

\end{itemize}

\section{MORE Dataset}
In this section, we provide a brief overview of the construction process for the MORE dataset, covering data collection and annotation, as well as a detailed analysis of the final dataset. Additional information on the dataset construction process can be found in Appendix \ref{dataset_construction}.

\subsection{Dataset Construction}

To construct the MORE dataset, we choose to use multimodal news data as a source rather than annotating existing MRE datasets primarily sourced from social media \cite{DBLP:conf/mm/ZhengFFCL021, DBLP:conf/icmcs/ZhengWFF021}. Multimodal news data has selective and well-edited images and textual titles, resulting in relatively good data quality, and often contains timely and informative knowledge. We obtained the data from The New York Times English news and Yahoo News from 2019 to 2022, resulting in a candidate set of 15,000 multimodal news data instances covering various topics. We filtered out unqualified data and obtained a meticulously selected dataset for our research purposes. Then the candidate multimodal news was annotated in the following three distinct stages.

\textbf{Stage 1: Entity Identification and Object Detection.} We utilized the AllenNLP named entity recognition tool\footnote{\url{https://allenai.org/allennlp}} and the Yolo V5 object detection tool\footnote{\url{https://github.com/ultralytics/yolov5}} to identify the entities in textual news titles and the object areas in the corresponding news images. All extracted objects and entities were reviewed and corrected manually by our annotators. 

\begin{figure}[ht]
    \centering
    \includegraphics[width=0.75\linewidth]{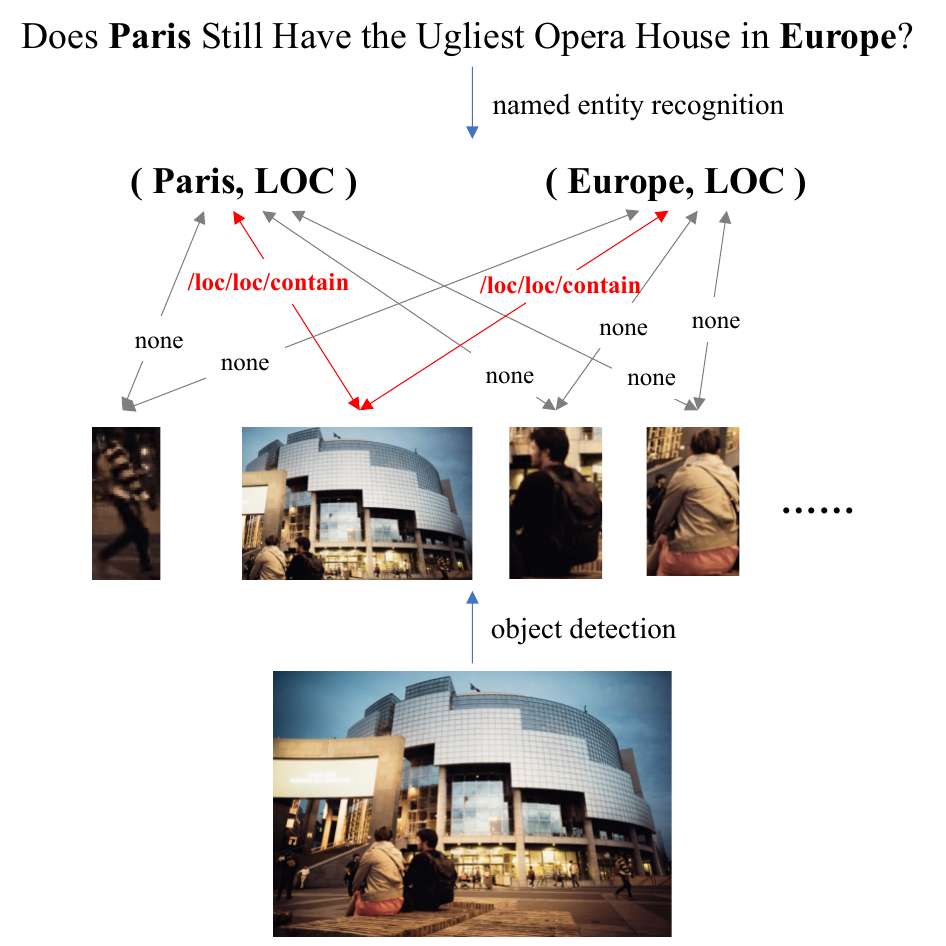}
    \caption{An example of annotation.}
    \label{fig:annotation}
\end{figure}

\textbf{Stage 2: Object-Entity Relation Annotation.} We recruited well-educated annotators to examine the textual titles and images and deduce the relations between the entities and objects. Relations were randomly assigned to annotators from the candidate set to ensure an unbiased annotation process. The data did not clearly indicate any pre-defined relations will be labeled as \verb|none| (Figure \ref{fig:annotation}). At least two annotators are required to independently review and annotate each data. In cases where there were discrepancies or conflicts in the annotations, a third annotator was consulted, and their decision was considered final. The weighted Cohen’s Kappa \cite{cohen1960coefficient} is used to measure the consistency between different annotators\footnote{The Kappa value is 0.7185 here.}.

\textbf{Stage 3: Object-Overlapped Data Filtering.} To refine the scope of multimodal object-entity relation extraction task, we only focused on relations in which visual objects did not co-occur with any entities mentioned in the textual news titles. This process filtered down the data from 15,000 to over 3,000 articles containing more than 20,000 object-entity relational facts. This approach ensured a dataset of only relatable object-entity relationships illustrated in images, rather than those that were already mentioned explicitly in the textual news titles, resulting in a more focused dataset for the task.

\subsection{Dataset Analysis}

This section provides a detailed analysis of MORE to facilitate a comprehensive understanding of the dataset and the task of Multimodal Object-Entity Relation Extraction. Through this analysis, we aim to elucidate the various facets of our dataset and the challenges associated with this task.

\textbf{Data Size.} Table \ref{table:dataset_comparison} presents a comprehensive comparison of MORE with several existing datasets for RE, including ACE 2003-2004 \cite{DBLP:conf/lrec/DoddingtonMPRSW04}, TACRED \cite{DBLP:conf/emnlp/ZhangZCAM17}, and the recently introduced multimodal relation extraction dataset MNRE \cite{DBLP:conf/mm/ZhengFFCL021}.

\begin{table}[ht]
\centering
\caption{Comparison of MORE with existing RE datasets.\newline (MM: multimodal, Img: images, Sent: sentences, VO: visual objects, Rel: relations)}
\label{table:dataset_comparison}
\begin{tabular}{ccccccc}
\hline
\textbf{Dataset} &\textbf{MM} & \textbf{Img} & \textbf{Sent} & \textbf{VO} & \textbf{Fact} &  \textbf{Rel} \\
\hline
ACE 03-04 \cite{DBLP:conf/lrec/DoddingtonMPRSW04} & No & - & 12,783 & - & 16,771  & 24 \\
TACRED \cite{DBLP:conf/emnlp/ZhangZCAM17} & No & - & 53,791 & - & 21,773 & 41\\

MNRE \cite{DBLP:conf/mm/ZhengFFCL021} & Yes & 9,201 & 9,201 & - &15,485 & 23 \\
\textbf{MORE} & Yes & 3,559 & 3,559 & 13,520 & 20,264 & 21\\
\hline
\end{tabular}
\end{table}

MORE consists of 21 distinct relation types and contains 20,264 multimodal relational facts annotated on 3,559 pairs of textual titles and corresponding images, making it one of the largest datasets for multimodal relation extraction in multiple aspects. Furthermore, this dataset includes 13,520 visual objects (3.8 objects per image on average), a crucial element that has been missing from previous datasets. 
We split the dataset into training, development, and testing sets consisting of 15,486, 1,742 and 3,036 facts respectively.

\begin{figure}[htbp]
    \centering
    \includegraphics[width=1.0\linewidth]{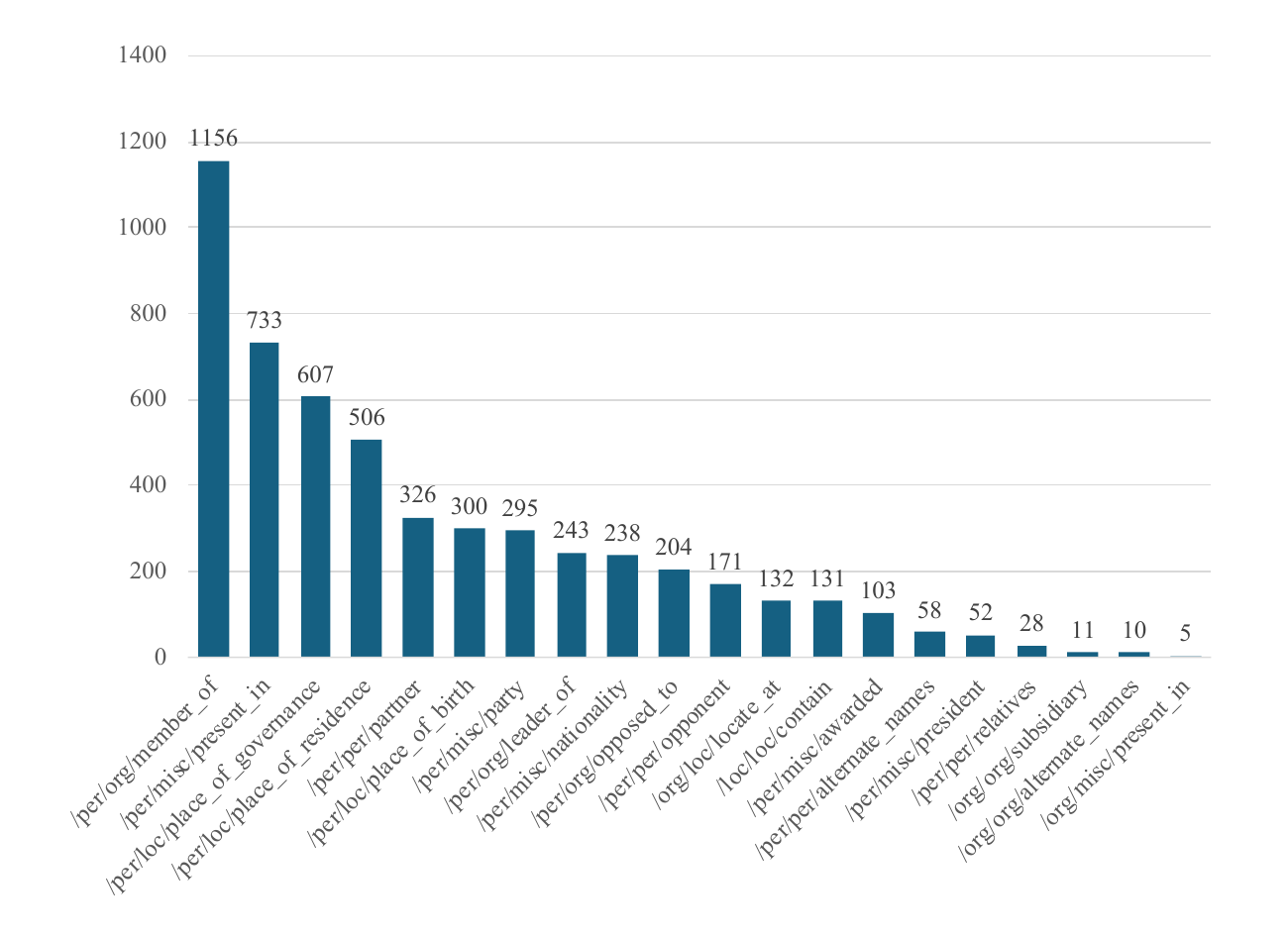}
    \caption{The distribution of relations.}
    \label{fig:relation_distribution}
\end{figure}

\textbf{Data Distribution.} There are 21 frequently-used relation types in the MORE dataset (Figure \ref{fig:relation_distribution}), which cover a broad range of personal life, location, etc. To gain a more in-depth understanding of the data, we conduct an analysis of the number of textual entities and visual objects present in each data instance. On average there are 1.5 entities per text and 3.8 objects per image. Table \ref{table:multiple_distribution} shows that only 791 instances (22.2\%) contain one entity in the text and one object in the image, while multiple entities or objects are present in the text or image in the remaining 77.8\% of data instances. This highlights the complexity of the task and underscores the need for models to possess more required capabilities.

\begin{table}[htbp]
    \centering
    \caption{
    Distributions of entities and visual objects.}
    \begin{tabular}{ccc}
    \hline
      & EntNum = 1 & EntNum > 1 \\
    \hline
    ObjNum = 1 & 791 (22.2\%) & 431 (12.1\%) \\
    ObjNum > 1 & 1,411 (39.7\%) & 926 (26\%) \\
    \hline
    \end{tabular}
    \label{table:multiple_distribution}
\end{table}

\textbf{Required Capabilities.} Conventional multimodal relation extraction tasks require a varying degree of understanding of the text and image modalities. Apart from these prerequisites, MORE demands the ability to address issues of semantic inconsistencies and perform \textit{multi-object disambiguation}. This involves identifying the relevant object in the image being referred to in the text. 

\begin{figure*}[htbp]
    \centering
    \includegraphics[width=0.9\linewidth]{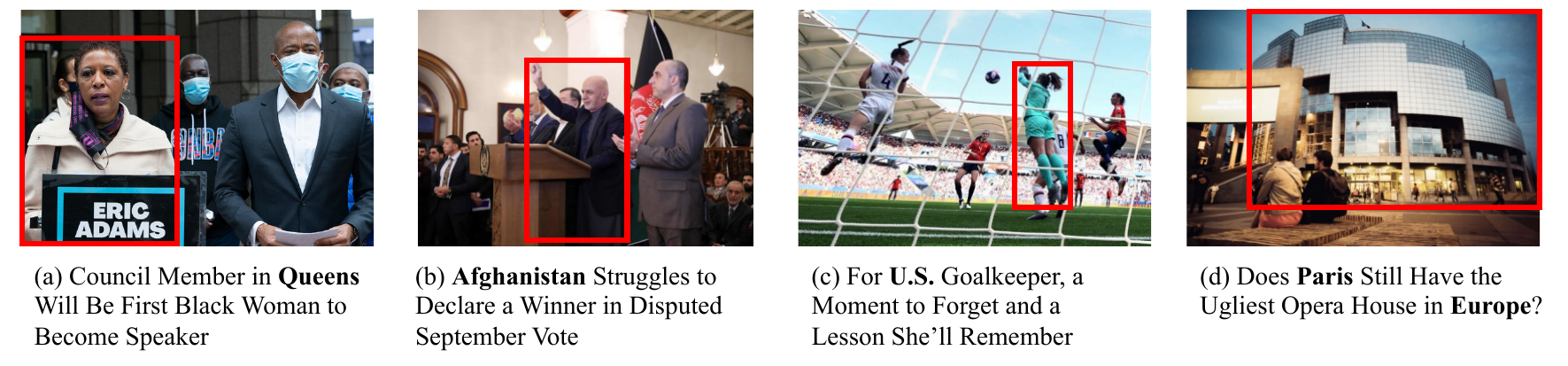}
    \caption{Examples for multi-object disambiguation. The red box identifies the correct object.}
    \label{fig:ability_example}
\end{figure*}

To further elaborate on the required capabilities of MORE, this study includes several examples from the dataset presented in Figure \ref{fig:ability_example}. Each image contains multiple objects that can be detected, but only part of them are referred to by the accompanying text. Compared to visual grounding, these examples pose unique challenges as the text does not directly describe the image as a caption and may include irrelevant information. Figure \ref{fig:ability_example}(a) demonstrates how aligning text that describes a "black woman" with the attributes of each object in the image can facilitate the identification of the target object. In this instance, the target object is the female on the left in the front row. In Figure \ref{fig:ability_example}(b), we illustrate how a comprehensive analysis of image information, such as object position, depth, and size, can aid in identifying the desired object when the text provides no clues. In Figure \ref{fig:ability_example}(c), understanding the term "goalkeeper" in the text can enable the identification of the second athlete on the right, wearing blue, as the target object in the image. Lastly, a more specific example is given in Figure \ref{fig:ability_example}(d), where multiple objects are present in the image, and the building is identified as the target object based on contextual information provided through the text rather than any person depicted in the image.

In summary, MORE requires models with a grasp of text and image modalities and the skill to identify relevant visual objects through text. This poses new challenges for the multimodal object-entity relation extraction task, as illustrated in Table \ref{table:ability}.

\begin{table}[htbp]
    \centering
    \caption{
    Abilities required in different RE tasks. \newline (IU: image understanding, TU: text understanding, MOD: multi-object disambiguation)
    }
    \begin{tabular}{c|ccc}
    \hline
     \textbf{Task} & \textbf{MOD} & \textbf{IU} & \textbf{TU} \\
    \hline
    Textual RE &  &  & \checkmark \\
    Multimodal RE & & \checkmark& \checkmark \\
    Object-Entity RE& \checkmark &\checkmark &\checkmark\\
    \hline
    \end{tabular}
    \label{table:ability}
\end{table}

\section{Experiments}
\subsection{Task Formulation}

In multimodal object-entity relation extraction, the objective is to predict relations between objects and entities based on both textual and image inputs. This task can be modeled using a function $F=(e,o,S,V) \to \mathcal{R}$, where $e$ and $o$ represent the pre-extracted textual entity and visual object, respectively. Given a sentence $S$ containing $e$ and the image $V$ containing $o$, the goal is to classify the corresponding relation tag $\mathcal{R}$ between $e$ and $o$.

\subsection{Baselines}

Multimodal object-entity relation extraction is a novel task that demands the prediction of relations using both textual and image object regions as inputs. Consequently, existing MRE methods have to undergo structural modifications to accommodate this task's requirements. Specifically, we concatenate the textual entity's embedding $t_h$ as the head embedding and the visual object's embedding $v_t$ as the tail embedding. Then we input them into the final relation classification head.

\begin{align}
    r = \mathrm{argmax}(\operatorname{MLP}([t_h, v_t]))
\end{align}
\noindent where $r \in \mathcal{R}$ represents the relation with the maximum prediction probability.

Based on the aforementioned modification, to explore the performance of existing methods on this task, we choose baseline models from two aspects: multimodal relation extraction (MRE) models and vision-language pre-training (VLP) models. 

\textbf{MRE Models}. \textbf{BERT+SG} \cite{DBLP:conf/icmcs/ZhengWFF021} concatenates textual representations from BERT with visual features generated by a scene graph tool \cite{DBLP:conf/cvpr/TangNHSZ20}. \textbf{Bert+SG+Att} \cite{DBLP:conf/icmcs/ZhengWFF021} leverages the attention mechanism to consider the semantic similarity between the visual graph and textual contents. \textbf{MEGA} \cite{DBLP:conf/mm/ZhengFFCL021} designs an efficient graph alignment that considers both structural similarity and semantic agreement between the visual and textual graphs. \textbf{IFAformer} \cite{DBLP:journals/corr/abs-2211-07504} is based on the Transformer and incorporates a prefix attention enhancement mechanism. \textbf{MKGformer} \cite{DBLP:conf/sigir/ChenZLDTXHSC22}, the current state-of-the-art for MRE, features a novel M-Encoder module designed to enable multi-level fusion of visual Transformer and text Transformer.

\textbf{VLP Models}. \textbf{ViLBERT} \cite{DBLP:journals/corr/abs-1908-03557} is an extension of BERT \cite{DBLP:conf/nips/VaswaniSPUJGKP17} to a multimodal two-stream model, allowing it to process both visual and textual inputs in separate streams through a co-attention Transformer layer. \textbf{VisualBERT} \cite{DBLP:conf/nips/LuBPL19} is a single-stream network that extends the Transformer model to adapt to visual inputs. These approaches demonstrate the ongoing effort in multimodal research to develop models that can effectively leverage and integrate information from multiple modalities.

\subsection{MOREformer}

\begin{figure*}[t]
    \centering
    \includegraphics[width=0.75\linewidth]{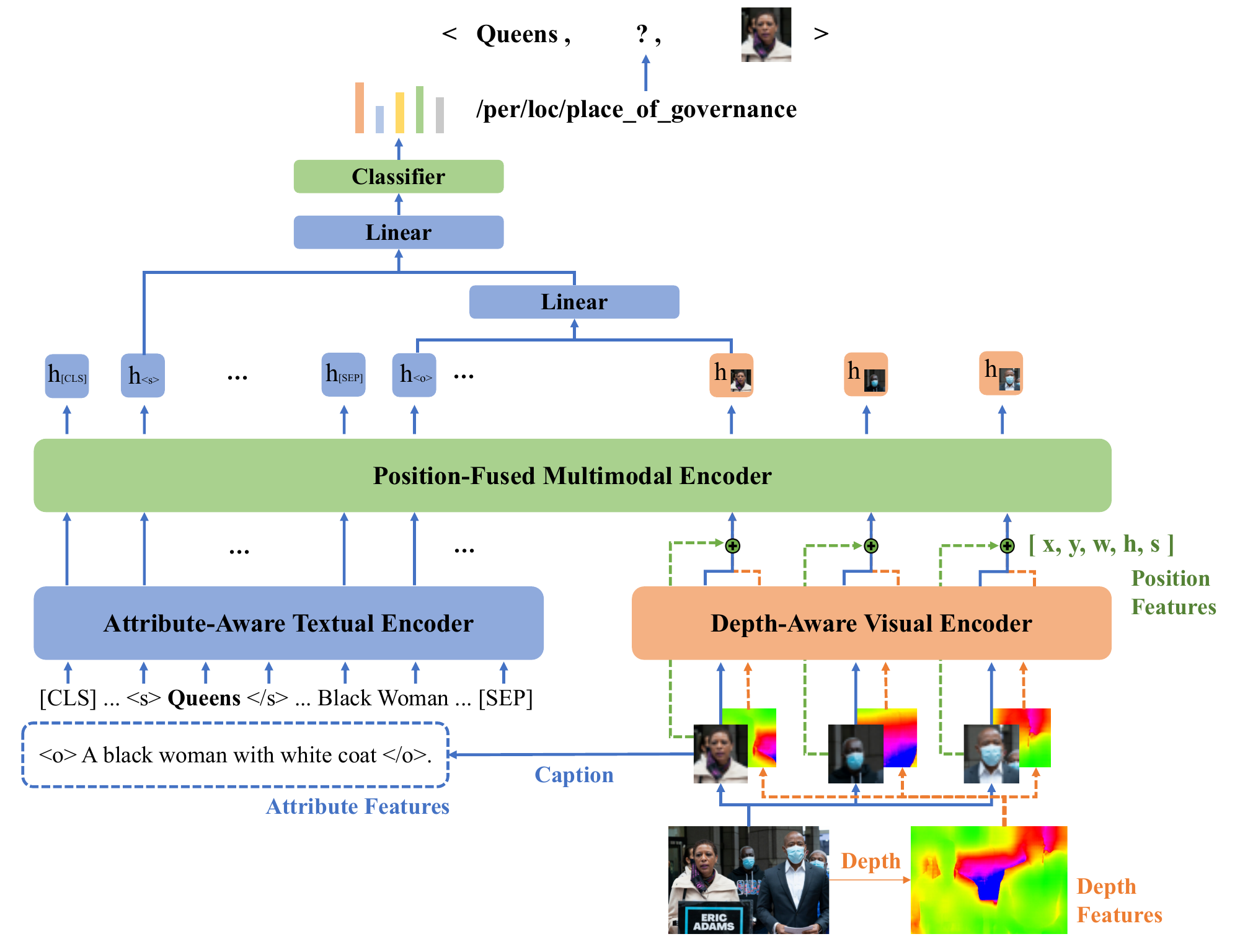}
    \caption{The framework of MOREformer.}
    \label{fig:framework}
\end{figure*}

Based on the MKGformer architecture, we introduce an enhanced approach called \textbf{MOREformer} (Figure \ref{fig:framework}). By leveraging attribute-aware textual encoding, depth-aware visual encoding, and position-fused multimodal encoding, our proposed model enhances its ability to perform multi-object disambiguation. 

\subsubsection{Recap of the MKGformer Architecture.} 

MKGformer utilizes the first $L_V$ layer of ViT \cite{DBLP:conf/iclr/DosovitskiyB0WZ21} as the visual encoder and the first $L_T$ layer of BERT\cite{DBLP:conf/nips/VaswaniSPUJGKP17} as the text encoder. In addition, MKGformer includes an M-encoder for the hierarchical interaction of multimodal information. The encoding strategies for the $\ell^{th}$ layer of M-Encoder are as follows:

\begin{equation}
    \begin{aligned}
    \bar{H}_{\ell}^{M_{t}},\bar{H}_{\ell}^{M_{v}} &= \mathrm{PGI}(H_{\ell-1}^{M_{t}},H_{\ell-1}^{M_{v}}), \;\;\ell=1 \dots L_{M} \\ 
     H_{\ell}^{M_{t}},H_{\ell}^{M_{v}} &= \mathrm{CAF}(\bar{H}_{\ell-1}^{M_{t}},\bar{H}_{\ell-1}^{M_{v}}), \;\;\ell=1 \dots L_{M}
    \end{aligned}
    \label{eq:M-Encoder}
\end{equation}

\noindent Where $H^{M_{t}}$ and $H^{M_{v}}$ represent text and visual representations, respectively. 

The Prefix Guided Interaction (PGI) module redefines the self-attention calculation of $\mathrm{head}^{M_t}$ and $\mathrm{head}^{M_v}$ for text and vision as follows:

\begin{equation}
    \begin{aligned}
         \mathrm{head}^{M_t} &= \mathrm{Attn}(x^{t}W_{q}^{t},x^{t}W_{k}^{t},x^{t}W_{v}^{t}),\\ 
         \mathrm{head}^{M_{v}} &= \mathrm{Attn}(x^{v}W_{q}^{v},\,[x^{v}W_{k}^{v},\,{x}^{t}\,W_{k}^{t}\,],\,[x^{v}W_{v}^{v},x^{t}\,W_{v}^{t}\,]) 
    \end{aligned}
\end{equation}

\noindent 
Where $[,]$ represents vector concatenate. 

The Correlation-Aware Fusion (CAF) module performs token-wise cross-modal interaction. The definition of CAF is as follows:

\begin{equation}
    \begin{aligned}
            Sim &= x^{t}(x^{v})^T \\
            \mathrm{Agg}_{i}(x^{v}) &= \mathrm{softmax}(Sim_{i})x^{v},(1\leq i<n) \\
            \mathrm{Agg}(x^{v}) &= \left[\mathrm{Agg}_{1}(x^{v});...,\mathrm{Agg}_{n}(x^{v})\right] \\
            \mathrm{FFN}(x^{t}) &= \mathrm{ReLU}(x^{t}W_{1}+\mathrm{b}_{1}+\mathrm{Agg}(x^{v})W_{3})\mathrm{W}_{2}+\mathrm{b}_{2}
    \end{aligned}
\end{equation}

\noindent 
Where $Sim$ represents the token-wise similarity between text and images. $x^{t}$ and $x^{v}$ represent textual and visual vectors, respectively. $W_*$ and $b_*$ are trainable parameters. 

\subsubsection{Attribute-Aware Textual Encoder}

The input text may comprise description information about the object. Therefore, modeling the relationship between the attribute and the description of the object in the text is beneficial to determine the object. We utilize the popular image caption model ClipCap\cite{DBLP:journals/corr/abs-2111-09734} to get attribute information about objects. We wrap the obtained object caption $S_{cap}$ with token $\texttt{<o>}$, and warp the corresponding entity in $S$ with the token $\texttt{<s>}$, and then stitch them with $\texttt{[SEP]}$.
The above sentences will serve as input for the text encoder.  In accordance with MKGformer, we use BERT's first $L_T$ layer as the base text encoder to generate the contextual representation $H^{T} \in  \mathbb{R}^{n \times d_t}$($n$ donates the length of the sequence) as follows:

\begin{equation}
    \begin{aligned}
    W^{cap} & = \mathrm{ClipCap}(o_k) \\
    W^T &= \{x_1, \dots, \texttt{<s>}, x_{i1}, \dots, x_{i|x_i|}, \texttt{</s>}, \dots, x_n\} \\
    \hat{W}^T & = \{\texttt{[CLS]}, W^T, \texttt{[SEP]}, \texttt{<o>}, W^{cap}, \texttt{</o> }\} \\
    E^{T} & = E_{\hat{W}^T} + E_{pos} + E_{seg} \\
    \bar{H}_{\ell}^{T} & =\operatorname{LN}\left(\operatorname{MHA}\left(H_{\ell-1}^{T}\right)\right)+H_{\ell-1}^{T} \;\; \ell=1 \ldots L_{t} \\
    H_{\ell}^{T} & =\operatorname{LN}\left(\operatorname{FFN}\left(\bar{H}_{\ell}^{T}\right)\right)+\bar{H}_{\ell}^{T} \; \ell=1 \ldots L_{T}
    \end{aligned}
    \label{eq: textual encoder}
\end{equation}

\noindent 
Where $o_k$ represents the object to be predicted, $x_i$ represents the $i^{th}$ entity in $S$. $\mathrm{ClipCap}$ is the image caption model. $E_{pos}$ and $E_{seg}$ are consistent with BERT. 
$\operatorname{MHA}$, $\operatorname{FFN}$, and $\operatorname{LN}$ represent the multi-head attention, feed-forward network, and Layernorm operator, respectively.

\subsubsection{Depth-Aware Visual Encoder}

The depth map contains the hierarchical information of the object, which is important for multi-object disambiguation. We use the depth estimation model S2R-DepthNet\cite{DBLP:conf/cvpr/ChenWCZ21} to obtain the depth map of the entire picture. Then we obtain the 
separate depth image from the area corresponding to each object, denoted as D-image. 
Following MKGformer, we employ the first $L_V$ layers of ViT as the visual encoder to extract image features. 
Given objects $o_k$($1 \leq k \leq m$)\footnote{Unless otherwise specified, the range of $k$ is $[1,m]$.}, we rescale the corresponding RGB-image $I^{RGB}_k $ and D-image $I^{D}_k$ to unified $H \times W$ pixels. $I^{RGB}_k \in \mathbb{R}^{3 \times H \times W}$ and $I^{D}_k \in \mathbb{R}^{1 \times H \times W}$ are then converted into RGB patch embedding $E^{rgb}_k \in \mathbb{R}^{u\times d_v} $ and Depth patch embedding $E^{d}_k \in \mathbb{R}^{u\times d_v}$ using the naive ViT patch embedding Layer, where $u = \frac{H \times W}{P^2}$, $P$ is pre-defined patch size and $d_v$ represents the dimension of hidden states of ViT. The position embedding $ E_{pos} \in \mathbb{R}^{ u \times d_v} $ is added to RGB patch embedding and depth patch embedding according to equation \eqref{eq:visual_encoder_1}:
\begin{equation}
\label{eq:visual_encoder_1}
    \begin{aligned}
\hat{E}_{k}^{V_{RGB}} &= E^{rgb}_k+E_{pos}\\
\hat{E}_{k}^{V_D} &= E^{d}_k+E_{pos}\\
    \end{aligned}
\end{equation}

RGB patch embedding and Depth patch embedding of all objects are concatenated to produce visual sequence patch embedding $H_{0}^V \in \mathbb{R}^{2 \times m \times u \times d_v} $. The visual representation is calculated by visual sequence patch embedding according to equation \eqref{eq:visual_encoder_2}:

\begin{equation}
\label{eq:visual_encoder_2}
    \begin{aligned}
    H_{0}^V = [\hat{E}_1^{V_{RGB}}, \hat{E}_1^{V_{D}}, \hat{E}_2^{V_{RGB}}, \hat{E}_2^{V_{D}}, ... , \hat{E}_m^{V_{RGB}}, \hat{E}_m^{V_{D}}] \\
    \bar{H}_{\ell}^{V}=\operatorname{MHA}\left(\operatorname{LN}\left(H_{\ell-1}^{V}\right)\right)+H_{\ell-1}^{V}, \;\; \ell=1 \ldots L_{V} \\    H_{\ell}^{V}=\operatorname{FFN}\left(\operatorname{LN}\left(\bar{H}_{\ell}^{V}\right)\right)+\bar{H}_{\ell}^{V}, \;\; \ell=1 \ldots L_{V}
    \end{aligned}
\end{equation}

\subsubsection{Position-Fused Multimodal Encoder}

Position commonly indicates the significance of objects. E.g., objects in the center or with a larger surface area may be more crucial.
We design a position-fused encoder to integrate object position information. Specifically,
we first conduct an average pooling operation on the visual representation of each object obtained by the Depth-Aware Visual Encoder to fuse the RGB representation and Depth representation of object $o_k$ as equation \eqref{eq:PEME}:

\begin{equation}
    \label{eq:PEME}
    \begin{aligned}
        \Tilde{H}^{V}_{L_V,k} & = \operatorname{AvgPool}(H_{L_{V},k}^{V})
    \end{aligned}
\end{equation}
where $H_{L_V,k}^{V} \in \mathbb{R}^{2 \times u \times d_v} $ is the visual representation of object $o_k$ in $H_{L_V}^{V}$ from equation \eqref{eq:visual_encoder_2}, and $\Tilde{H}^{V}_{L_V,k}$ is a $d_v$-dimensional embedding, which used as the RGB-D visual feature of the object $o_k$.

To model the importance of the object, we fuse the RGB-D visual feature and the position feature of each object. The position feature of the object $o_k$ can be represented by a 5-dimensional vector: 
\begin{align}
    P^{loc}_k & = [ x_{center}, y_{center}, w_{box}, h_{box}, s_{box} ]
\end{align}
where $w_{box}$, $h_{box}$, and $s_{box}$ are the width, height, and area of $o_k$, respectively. We normalize them into the range $[0,1]$. Then the position feature $P^{loc}_k $ is fused with the object's RGB-D visual feature $\Tilde{H}_{L_V,k}^{V}$. The input for the multimodal encoder is as follows:
\begin{equation}
    \begin{aligned}
        H_{0}^{M_{t}} & = H_{L_{T}}^{T} \\
        H_{0}^{M_{v}} & = \Tilde{H}_{L_{V}}^{V} + (P^{loc})W^{loc} + b^{loc} 
    \end{aligned}
\end{equation}
In which $H_{L_{T}}^{T} \in \mathbb{R}^{n \times d_t} $ is from equation \eqref{eq: textual encoder} and $\Tilde{H}_{L_{V}}^{V} \in \mathbb{R}^{m \times d_v}$ is derived from \eqref{eq:PEME}.

We follow equation \eqref{eq:M-Encoder} to model the multimodal features across the last $L_M$ layers of ViT and BERT with multi-level fusion.
Finally, we concatenate the visual representation and the attribute start position indicator $\texttt{<o>}$ as object embedding, the entity's start position indicator $\texttt{<s>}$ is used as entity embedding, then object embedding and entity embedding are concatenated to predict the relation, which is illustrated as Equation \eqref{eq:PEME2}:
\begin{equation}
    \label{eq:PEME2}
    \begin{aligned}
        m_k &= [ h^{M_{v}}_k , h^{M_{t}}_{<o>} ] W^{m} + b^{m} \\
        r &= \mathrm{argmax} (\operatorname{MLP}([h^{M_{t}}_{<s>}, m_k]))
    \end{aligned}
\end{equation}
where $h^{M_{t}}_{<s>}$ and $h^{M_{t}}_{<o>}$ is the multimodal representation of token $\texttt{<s>}$ and $\texttt{<o>}$ from $H_{L_M}^{M_{t}}$, $h^{M_{v}}_{k}$ is the multimodal representation of object $o_k$ from $H_{L_M}^{M_{v}}$ .


\subsection{Experimental Setup}

We implement the MOREformer using the publicly available MKGformer \cite{DBLP:conf/sigir/ChenZLDTXHSC22} as a foundation. The textual semantic representation is acquired through initializing the textual representation with pre-trained BERT, with a dimension of 768. Additionally, the dimension of visual object features extracted from the image is set at 4096. The maximum numbers of token sequences and objects are limited to 96 and 10, respectively. Our model is optimized using the AdamW optimizer \cite{DBLP:conf/iclr/LoshchilovH19}, with a base learning rate set at 1e-5 and a batch size of 32. In our experiment, we employ a dropout rate of 0.5. It should be noted that, in line with the MKGformer, the layer of M-Encoder is set at $L_M = 3$, and we experiment with BERT\_base and ViT-B/32 for all experiments. And following the conventional MRE task, we utilize accuracy, precision, recall, and F1 value as the evaluation metrics. Since the MORE dataset has an imbalanced label distribution, we choose F1 score as the main evaluation metric for measuring the performance of a class-imbalanced task.

\section{Result and Analysis}

\subsection{Overall Results}

We conduct experiments on the MORE dataset, and Table \ref{table:results} illustrates the overall results on the test set of MORE. The following observations are made: 1) BERT+SG, BERT+SG+Att, and MEGA attempt to map from visual to textual contents based on visual scene graph. However, MORE dataset is compiled from real-world news articles where textual descriptions may not always directly correspond to accompanying images. This generates inconsistencies in semantics between textual and visual data, leading to noise derived from the scene graph, resulting in poor performance of these models on the dataset. 2) IFAformer and MKGformer, being MRE models, take advantage of fine-grained image-text alignment to alleviate errors caused by irrelevant visual information. Compared to other related models, their performance on the dataset has substantially improved. 3) VLP models mainly learn the semantic correspondence between different modalities by pre-training on large-scale data. It is evident that pre-training facilitates the learning of fine-grained semantic alignment between visual and textual modalities, hence significantly enhancing the performance on this task. 4) Our proposed MOREformer, which further integrates information such as visual object attributes, depth, and position, enhances the model's ability to capture relations between image and text, and alleviate the multi-object ambiguity, leading to a significant improvement in performance. Our model outperforms MKGformer, which currently holds the highest performance among MRE models, by 8\%. Moreover, compared to ViLBERT, which currently has the best performance among VLP models, our model shows an increase of over 1.6\% in scores on this dataset.

\begin{table}[htbp]
\caption{
    The overall performance of MOREformer and other state-of-the-art methods on MORE}
\label{table:results}
\centering
    \begin{tabular}{c|cccc}
    \hline
    \textbf{Model} & \textbf{Accuracy} &\textbf{Precision}             &\textbf{Recall} & \textbf{F1-Score} \\
    \hline
        BERT+SG &61.79 & 29.61 & 41.27 & 34.48\\
        BERT+SG+Att  & 63.74 & 31.10 & 39.28 & 34.71 \\
        MEGA & 65.97 & 33.30 & 38.53 & 35.72 \\
        IFAformer  & 79.28 & 55.13 & 54.24 & 54.68 \\
        MKGformer & 80.17 & 55.76 & 53.74 & 54.73 \\
    \hline
        VisualBERT & 82.84 & 58.18 & 61.22 & 59.66 \\
        ViLBERT & \textbf{83.50} & \textbf{62.53} & 59.73 & 61.10 \\
    \hline
     \textbf{MOREformer} & \textbf{83.50} & 62.18 & \textbf{63.34} & \textbf{62.75} \\
    \hline
    \end{tabular}
\end{table}

\subsection{Ablation Study}

We conduct feature ablation studies on the MOREformer to investigate the contribution of different features in multimodal object-entity relation extraction. We systematically analyze the effect of visual object attribute, depth, and position on the model's performance. Table \ref{table:ablations} presents the results of the experiments, indicating that all the aforementioned features have a positive contribution to the model's performance. Furthermore, the findings reveal that integrating more features may lead to better performance, and the fusion of more features can almost achieve the best performance of the baseline model.

In addition, our analysis highlights that the position feature has the most pronounced impact on the model's performance, particularly in images containing multiple objects. A reasonable explanation is that as the primary objects often occupy a large area in the center of an image, accurate identification of the target object may be achieved solely by the position information. This finding aligns with the specific characteristics of the MORE dataset and provides important insights into improving multimodal object-entity relation extraction.

\begin{table}[htbp]
\caption{Feature ablations (P:position, A:attribute, D:depth)}
\label{table:ablations}
\centering
\begin{tabular}{ccc|cc}
\hline
 \multicolumn{3}{c|}{\textbf{Feature}}  & \textbf{Accuracy} & \textbf{F1-Score} \\
\hline
 P& A& D & & \\
\hline
 &  &  & 80.17  & 54.73 \\
\checkmark & &  & 83.07 & 60.58\\
& \checkmark &  & 80.11 & 56.95\\
& & \checkmark &  80.70  & 56.55 \\
\checkmark & \checkmark &  & 82.87 & 61.91 \\
\checkmark & & \checkmark & 82.64 & 59.63  \\
& \checkmark & \checkmark & 81.03 & 58.95  \\
\hline
\checkmark& \checkmark& \checkmark & 83.50 & 62.75 \\
\hline
\end{tabular}
\end{table}

\subsection{Detailed Analysis}

This section offers a meticulous evaluation of the model's performance, assessing its ability to handle multiple entities and objects, its effectiveness in handling long-tail relations, and its ability to identify mentioned objects within the text.

\subsubsection{Performance on multiple entities/objects}

We evaluate the performance of the model on different combinations of multiple entities in text and multiple objects in images. Table \ref{table:category} illustrates that the model's performance is notably high when there is only one entity in the text and one object in the image. However, as the number of entities or objects increases, the performance of the model decreases significantly. Additionally, the model's performance drops sharply when there are multiple entities in the text and multiple objects in the image. This indicates that an increase in the number of entities and objects poses greater difficulties for the model in predicting relations. It also demonstrates that the MORE dataset is highly challenging.

Notably, when there are multiple objects in an image, the F1 score of the model decreases significantly, while the accuracy remains high. This could be due to the dataset being class-imbalanced, with a considerable number of \verb|none| relations, resulting in the model accurately predicting these relations but with less practical meaning. Therefore, the model needs to overcome the challenges of overfitting on \verb|none| relations when dealing with instances containing multiple entities and objects to achieve better performance.

\begin{table}[h]
\caption{The performance on multiple entities/objects. (Ent: the number of entities in a text, Obj: the number of objects in an image)}
\label{table:category}
\centering
\begin{tabular}{c|cccc}
\hline
  & \textbf{Ent=1} & \textbf{Ent=1} & \textbf{Ent>1} & \textbf{Ent>1} \\
  & \textbf{Obj=1} & \textbf{Obj>1} & \textbf{Obj=1} & \textbf{Obj>1} \\
\hline
\textbf{none} Ratio & 5.00\% & 74.10\% & 33.11\% & 80.90\% \\
\hline
Accuracy & 80.00 & 84.57 & 67.55 & 84.43\\
Precision & 82.98 & 63.18 & 66.36 & 53.63 \\
Recall & 82.11 &  66.31 & 72.28 & 52.47 \\
F1-Score & 82.54 & 64.71 & 69.19 & 53.04 \\
\hline
\end{tabular}
\end{table}

\subsubsection{Performance on long-tail relations}


In real-world situations, data often exhibits an unbalanced distribution, resulting in a long-tail phenomenon. This is also evident in the MORE dataset, where relationships have an evident long-tail distribution. Consequently, models may have a tendency to fit a large proportion of the relationship data, while disregarding the long-tail relations, leading to high F1 but low macro-F1 values. We evaluate the performance of our model using the macro-F1 metric on long-tail relations, which provides a more comprehensive assessment of the model's effectiveness in predicting all relationship types. The evaluation results are presented in Table \ref{table:longtail}. The findings highlight that, despite our model outperforming the baseline models, all models display unsatisfactory performance in addressing long-tail relations. Therefore, there is an urgent need for more effective techniques to mitigate the long-tail problem in this task.

\begin{table}[h]
\caption{Macro-F1 on MORE dataset}
\label{table:longtail}
\centering
    \begin{tabular}{cccc}
    \hline
     \textbf{Model Name}  &\textbf{Precision} &\textbf{Recall} & \textbf{Macro F1-Score} \\
    \hline
     MKGformer & 48.50 & 42.60 & 43.57 \\
     ViLBERT  & 49.13 & 49.37 & 48.66 \\
    \hline
    \textbf{MOREformer} & \textbf{51.02} & \textbf{50.66} & \textbf{50.02} \\
    \hline
    \end{tabular}
\end{table}

\subsubsection{Identify correct objects mentioned in the text}

\begin{figure}[htbp]
    \centering
    \includegraphics[width=\linewidth]{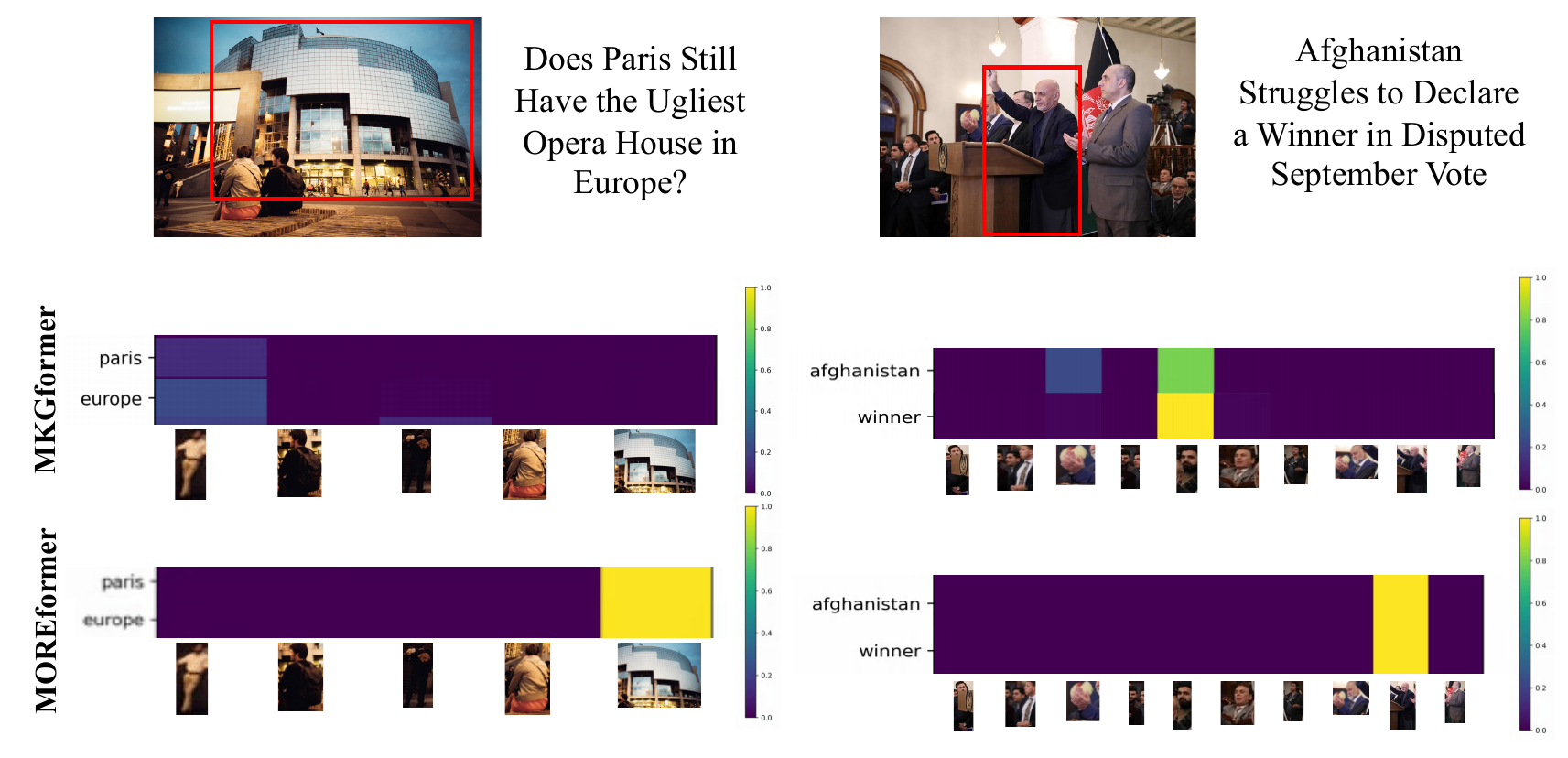}
    \caption{Visualization analysis.}
    \label{fig:visualization}
\end{figure}

In the MORE dataset, each image contains 3.8 objects on average, with the maximum number detected being more than 10. 
Consequently, disambiguating between multiple objects presents a significant challenge. The weak correlation between text and image further complicates multi-object disambiguation. We conduct this experiment to show the challenge of the Multimodal Object-Entity Relation Extraction task. More specifically, it is difficult to find the correct object corresponding to the entity in text when an image contains multiple objects.
We have evaluated the performance of our model and two well-performed baseline models on multi-object disambiguation in the MORE dataset. Specifically, a predicted triple is deemed correct on condition that: 1) the entity and object of the predicted triple are correct, and 2) the predicted relation and the golden relation are not \verb|none|. The results, as presented in Table \ref{table:mod}, demonstrate that our model significantly outperforms the baseline models, especially on the subset of instances that have more than 1 object. This suggests that by leveraging more feature information, the model can better identify the visual objects referred to in the text. Moreover, we have conducted a visualization analysis to examine whether the model's attention is primarily focused on the correct area of the image during prediction. Our visualization results, illustrated in Figure \ref{fig:visualization}, indicate that our model is better at identifying the relevant objects in images with multiple objects, thereby enhancing the accuracy of predicting relational facts.

\begin{table}[h]
\caption{Multi-object disambiguation analysis\newline (full dataset / subset of object number > 1)}
\label{table:mod}
\centering
    \begin{tabular}{cccc}
    \hline
     \textbf{Model Name} &\textbf{Precision} &\textbf{Recall} & \textbf{F1-Score} \\
    \hline
     MKGformer & 70.12/66.08 & 67.58/62.29 & 68.83/64.13 \\
     ViLBERT  & \textbf{75.31}/\textbf{72.30} & 75.31/70.41 & 75.31/71.34 \\
    \hline
    \textbf{MOREformer} & 72.28/71.43 & \textbf{79.75}/\textbf{76.61} & \textbf{75.84}/\textbf{73.93} \\
    \hline
    \end{tabular}
\end{table}

\section{Related Works}

Numerous efforts have been devoted to extracting relational facts from multimodal data and developing related datasets. Notably, MNRE \cite{DBLP:conf/mm/ZhengFFCL021, DBLP:conf/icmcs/ZhengWFF021} is the first multimodal relation extraction dataset, with a series of work demonstrating that incorporating multimodal information and efficient alignment strategy for textual and visual representations can effectively boost the performance of relation extraction in social media texts \cite{DBLP:conf/icmcs/WangYLTJYZX22, zhao2023tsvfn, DBLP:conf/emnlp/WangCJXTL22, DBLP:conf/coling/0023HDWSSX22}. Wan et al. \cite{DBLP:conf/aaai/WanZDHYP21} presented multimodal social relation datasets and proposed a few-shot learning based approach to extracting social relations from both texts and face images. Chen et al. \cite{DBLP:conf/sigir/ChenZLDTXHSC22} leveraged a hybrid transformer architecture with unified input-output for diverse multimodal tasks. Another crucial aspect of related research is vision-language pre-training models. Several studies have explored the proper designs of objects and model architectures to enable the model to identify associations between different modalities. Representative works in this area include VisualBERT \cite{DBLP:journals/corr/abs-1908-03557}, 
ViLBERT \cite{DBLP:conf/nips/LuBPL19}, CLIP-ViL \cite{DBLP:conf/iclr/ShenLTBRCYK22}, METER \cite{DBLP:conf/cvpr/DouXGWWWZZYP0022} and X-VLM \cite{DBLP:conf/icml/ZengZL22}. These works demonstrate the importance and effectiveness of utilizing multimodal information for relation extraction and highlight the potential of pre-trained multimodal models for enhancing performance.

\section{Conclusion}
In order to broaden the relational facts extraction from multimodal data, we propose a new task, namely Multimodal Object-Entity Relation Extraction, and correspondingly construct a new high-quality large-scale multimodal relational facts extraction dataset named MORE for this task. This dataset provides new insights for tackling the problems of semantic inconsistencies and multi-object ambiguity by leveraging both textual and visual information. Additionally, we compare and analyze several state-of-the-art multimodal relation extraction baselines. Our results demonstrate that these baselines suffer performance decline when tested on the MORE dataset, and that a proper way of incorporating textual and visual information can improve the multimodal object-entity relation extraction performance. In future work, we will explore ways to enhance the discriminative ability of learned representations \cite{liang2023knowledge} and optimally fuse the complementary information of each modal to uncover the intrinsic structure of multimodal relational facts \cite{DBLP:conf/mm/WanLLLWZ22, wan2023auto}.

\begin{acks}
We would like to express our gratitude to the anonymous reviewers for their constructive comments, as well as to all the hardworking and professionally dedicated annotators. This work is supported by the National Natural Science Foundation of China (project Nos. 62176115, 61936012, 62206126, and 61976114).
\end{acks}

\bibliographystyle{ACM-Reference-Format}
\bibliography{samples}


\begin{thebibliography}{35}


\ifx \showCODEN    \undefined \def \showCODEN     #1{\unskip}     \fi
\ifx \showDOI      \undefined \def \showDOI       #1{#1}\fi
\ifx \showISBNx    \undefined \def \showISBNx     #1{\unskip}     \fi
\ifx \showISBNxiii \undefined \def \showISBNxiii  #1{\unskip}     \fi
\ifx \showISSN     \undefined \def \showISSN      #1{\unskip}     \fi
\ifx \showLCCN     \undefined \def \showLCCN      #1{\unskip}     \fi
\ifx \shownote     \undefined \def \shownote      #1{#1}          \fi
\ifx \showarticletitle \undefined \def \showarticletitle #1{#1}   \fi
\ifx \showURL      \undefined \def \showURL       {\relax}        \fi
\providecommand\bibfield[2]{#2}
\providecommand\bibinfo[2]{#2}
\providecommand\natexlab[1]{#1}
\providecommand\showeprint[2][]{arXiv:#2}

\bibitem[Chen et~al\mbox{.}(2021)]%
        {DBLP:conf/cvpr/ChenWCZ21}
\bibfield{author}{\bibinfo{person}{Xiaotian Chen}, \bibinfo{person}{Yuwang Wang}, \bibinfo{person}{Xuejin Chen}, {and} \bibinfo{person}{Wenjun Zeng}.} \bibinfo{year}{2021}\natexlab{}.
\newblock \showarticletitle{S2R-DepthNet: Learning a Generalizable Depth-Specific Structural Representation}. In \bibinfo{booktitle}{\emph{{IEEE} Conference on Computer Vision and Pattern Recognition, {CVPR} 2021, virtual, June 19-25, 2021}}. \bibinfo{publisher}{Computer Vision Foundation / {IEEE}}, \bibinfo{pages}{3034--3043}.
\newblock
\urldef\tempurl%
\url{https://doi.org/10.1109/CVPR46437.2021.00305}
\showDOI{\tempurl}


\bibitem[Chen et~al\mbox{.}(2022)]%
        {DBLP:conf/sigir/ChenZLDTXHSC22}
\bibfield{author}{\bibinfo{person}{Xiang Chen}, \bibinfo{person}{Ningyu Zhang}, \bibinfo{person}{Lei Li}, \bibinfo{person}{Shumin Deng}, \bibinfo{person}{Chuanqi Tan}, \bibinfo{person}{Changliang Xu}, \bibinfo{person}{Fei Huang}, \bibinfo{person}{Luo Si}, {and} \bibinfo{person}{Huajun Chen}.} \bibinfo{year}{2022}\natexlab{}.
\newblock \showarticletitle{Hybrid Transformer with Multi-level Fusion for Multimodal Knowledge Graph Completion}. In \bibinfo{booktitle}{\emph{{SIGIR} '22: The 45th International {ACM} {SIGIR} Conference on Research and Development in Information Retrieval, Madrid, Spain, July 11 - 15, 2022}}, \bibfield{editor}{\bibinfo{person}{Enrique Amig{\'{o}}}, \bibinfo{person}{Pablo Castells}, \bibinfo{person}{Julio Gonzalo}, \bibinfo{person}{Ben Carterette}, \bibinfo{person}{J.~Shane Culpepper}, {and} \bibinfo{person}{Gabriella Kazai}} (Eds.). \bibinfo{publisher}{{ACM}}, \bibinfo{pages}{904--915}.
\newblock
\urldef\tempurl%
\url{https://doi.org/10.1145/3477495.3531992}
\showDOI{\tempurl}


\bibitem[Cohen(1960)]%
        {cohen1960coefficient}
\bibfield{author}{\bibinfo{person}{Jacob Cohen}.} \bibinfo{year}{1960}\natexlab{}.
\newblock \showarticletitle{A coefficient of agreement for nominal scales}.
\newblock \bibinfo{journal}{\emph{Educational and psychological measurement}} \bibinfo{volume}{20}, \bibinfo{number}{1} (\bibinfo{year}{1960}), \bibinfo{pages}{37--46}.
\newblock


\bibitem[Doddington et~al\mbox{.}(2004)]%
        {DBLP:conf/lrec/DoddingtonMPRSW04}
\bibfield{author}{\bibinfo{person}{George~R. Doddington}, \bibinfo{person}{Alexis Mitchell}, \bibinfo{person}{Mark~A. Przybocki}, \bibinfo{person}{Lance~A. Ramshaw}, \bibinfo{person}{Stephanie~M. Strassel}, {and} \bibinfo{person}{Ralph~M. Weischedel}.} \bibinfo{year}{2004}\natexlab{}.
\newblock \showarticletitle{The Automatic Content Extraction {(ACE)} Program - Tasks, Data, and Evaluation}. In \bibinfo{booktitle}{\emph{Proceedings of the Fourth International Conference on Language Resources and Evaluation, {LREC} 2004, May 26-28, 2004, Lisbon, Portugal}}. \bibinfo{publisher}{European Language Resources Association}.
\newblock
\urldef\tempurl%
\url{http://www.lrec-conf.org/proceedings/lrec2004/summaries/5.htm}
\showURL{%
\tempurl}


\bibitem[Dosovitskiy et~al\mbox{.}(2021)]%
        {DBLP:conf/iclr/DosovitskiyB0WZ21}
\bibfield{author}{\bibinfo{person}{Alexey Dosovitskiy}, \bibinfo{person}{Lucas Beyer}, \bibinfo{person}{Alexander Kolesnikov}, \bibinfo{person}{Dirk Weissenborn}, \bibinfo{person}{Xiaohua Zhai}, \bibinfo{person}{Thomas Unterthiner}, \bibinfo{person}{Mostafa Dehghani}, \bibinfo{person}{Matthias Minderer}, \bibinfo{person}{Georg Heigold}, \bibinfo{person}{Sylvain Gelly}, \bibinfo{person}{Jakob Uszkoreit}, {and} \bibinfo{person}{Neil Houlsby}.} \bibinfo{year}{2021}\natexlab{}.
\newblock \showarticletitle{An Image is Worth 16x16 Words: Transformers for Image Recognition at Scale}. In \bibinfo{booktitle}{\emph{9th International Conference on Learning Representations, {ICLR} 2021, Virtual Event, Austria, May 3-7, 2021}}. \bibinfo{publisher}{OpenReview.net}.
\newblock
\urldef\tempurl%
\url{https://openreview.net/forum?id=YicbFdNTTy}
\showURL{%
\tempurl}


\bibitem[Dou et~al\mbox{.}(2022)]%
        {DBLP:conf/cvpr/DouXGWWWZZYP0022}
\bibfield{author}{\bibinfo{person}{Zi{-}Yi Dou}, \bibinfo{person}{Yichong Xu}, \bibinfo{person}{Zhe Gan}, \bibinfo{person}{Jianfeng Wang}, \bibinfo{person}{Shuohang Wang}, \bibinfo{person}{Lijuan Wang}, \bibinfo{person}{Chenguang Zhu}, \bibinfo{person}{Pengchuan Zhang}, \bibinfo{person}{Lu Yuan}, \bibinfo{person}{Nanyun Peng}, \bibinfo{person}{Zicheng Liu}, {and} \bibinfo{person}{Michael Zeng}.} \bibinfo{year}{2022}\natexlab{}.
\newblock \showarticletitle{An Empirical Study of Training End-to-End Vision-and-Language Transformers}. In \bibinfo{booktitle}{\emph{{IEEE/CVF} Conference on Computer Vision and Pattern Recognition, {CVPR} 2022, New Orleans, LA, USA, June 18-24, 2022}}. \bibinfo{publisher}{{IEEE}}, \bibinfo{pages}{18145--18155}.
\newblock
\urldef\tempurl%
\url{https://doi.org/10.1109/CVPR52688.2022.01763}
\showDOI{\tempurl}


\bibitem[Ferrada et~al\mbox{.}(2017a)]%
        {DBLP:conf/semweb/FerradaBH17a}
\bibfield{author}{\bibinfo{person}{Sebasti{\'{a}}n Ferrada}, \bibinfo{person}{Benjamin Bustos}, {and} \bibinfo{person}{Aidan Hogan}.} \bibinfo{year}{2017}\natexlab{a}.
\newblock \showarticletitle{Answering Visuo-Semantic Queries with IMGpedia}. In \bibinfo{booktitle}{\emph{Proceedings of the {ISWC} 2017 Posters {\&} Demonstrations and Industry Tracks co-located with 16th International Semantic Web Conference {(ISWC} 2017), Vienna, Austria, October 23rd - to - 25th, 2017}} \emph{(\bibinfo{series}{{CEUR} Workshop Proceedings}, Vol.~\bibinfo{volume}{1963})}, \bibfield{editor}{\bibinfo{person}{Nadeschda Nikitina}, \bibinfo{person}{Dezhao Song}, \bibinfo{person}{Achille Fokoue}, {and} \bibinfo{person}{Peter Haase}} (Eds.). \bibinfo{publisher}{CEUR-WS.org}.
\newblock
\urldef\tempurl%
\url{https://ceur-ws.org/Vol-1963/paper615.pdf}
\showURL{%
\tempurl}


\bibitem[Ferrada et~al\mbox{.}(2017b)]%
        {DBLP:conf/semweb/FerradaBH17}
\bibfield{author}{\bibinfo{person}{Sebasti{\'{a}}n Ferrada}, \bibinfo{person}{Benjamin Bustos}, {and} \bibinfo{person}{Aidan Hogan}.} \bibinfo{year}{2017}\natexlab{b}.
\newblock \showarticletitle{IMGpedia: {A} Linked Dataset with Content-Based Analysis of Wikimedia Images}. In \bibinfo{booktitle}{\emph{The Semantic Web - {ISWC} 2017 - 16th International Semantic Web Conference, Vienna, Austria, October 21-25, 2017, Proceedings, Part {II}}} \emph{(\bibinfo{series}{Lecture Notes in Computer Science}, Vol.~\bibinfo{volume}{10588})}, \bibfield{editor}{\bibinfo{person}{Claudia d'Amato}, \bibinfo{person}{Miriam Fern{\'{a}}ndez}, \bibinfo{person}{Valentina A.~M. Tamma}, \bibinfo{person}{Freddy L{\'{e}}cu{\'{e}}}, \bibinfo{person}{Philippe Cudr{\'{e}}{-}Mauroux}, \bibinfo{person}{Juan~F. Sequeda}, \bibinfo{person}{Christoph Lange}, {and} \bibinfo{person}{Jeff Heflin}} (Eds.). \bibinfo{publisher}{Springer}, \bibinfo{pages}{84--93}.
\newblock
\urldef\tempurl%
\url{https://doi.org/10.1007/978-3-319-68204-4\_8}
\showDOI{\tempurl}


\bibitem[Li et~al\mbox{.}(2022)]%
        {DBLP:journals/corr/abs-2211-07504}
\bibfield{author}{\bibinfo{person}{Lei Li}, \bibinfo{person}{Xiang Chen}, \bibinfo{person}{Shuofei Qiao}, \bibinfo{person}{Feiyu Xiong}, \bibinfo{person}{Huajun Chen}, {and} \bibinfo{person}{Ningyu Zhang}.} \bibinfo{year}{2022}\natexlab{}.
\newblock \showarticletitle{On Analyzing the Role of Image for Visual-enhanced Relation Extraction}.
\newblock \bibinfo{journal}{\emph{CoRR}}  \bibinfo{volume}{abs/2211.07504} (\bibinfo{year}{2022}).
\newblock
\urldef\tempurl%
\url{https://doi.org/10.48550/arXiv.2211.07504}
\showDOI{\tempurl}
\showeprint[arXiv]{2211.07504}


\bibitem[Li et~al\mbox{.}(2019)]%
        {DBLP:journals/corr/abs-1908-03557}
\bibfield{author}{\bibinfo{person}{Liunian~Harold Li}, \bibinfo{person}{Mark Yatskar}, \bibinfo{person}{Da Yin}, \bibinfo{person}{Cho{-}Jui Hsieh}, {and} \bibinfo{person}{Kai{-}Wei Chang}.} \bibinfo{year}{2019}\natexlab{}.
\newblock \showarticletitle{VisualBERT: {A} Simple and Performant Baseline for Vision and Language}.
\newblock \bibinfo{journal}{\emph{CoRR}}  \bibinfo{volume}{abs/1908.03557} (\bibinfo{year}{2019}).
\newblock
\showeprint[arXiv]{1908.03557}
\urldef\tempurl%
\url{http://arxiv.org/abs/1908.03557}
\showURL{%
\tempurl}


\bibitem[Liang et~al\mbox{.}(2023)]%
        {liang2023knowledge}
\bibfield{author}{\bibinfo{person}{Ke Liang}, \bibinfo{person}{Yue Liu}, \bibinfo{person}{Sihang Zhou}, \bibinfo{person}{Wenxuan Tu}, \bibinfo{person}{Yi Wen}, \bibinfo{person}{Xihong Yang}, \bibinfo{person}{Xiangjun Dong}, {and} \bibinfo{person}{Xinwang Liu}.} \bibinfo{year}{2023}\natexlab{}.
\newblock \showarticletitle{Knowledge Graph Contrastive Learning Based on Relation-Symmetrical Structure}.
\newblock \bibinfo{journal}{\emph{IEEE Transactions on Knowledge and Data Engineering}} (\bibinfo{year}{2023}).
\newblock


\bibitem[Liang et~al\mbox{.}(2022)]%
        {liang2022reasoning}
\bibfield{author}{\bibinfo{person}{Ke Liang}, \bibinfo{person}{Lingyuan Meng}, \bibinfo{person}{Meng Liu}, \bibinfo{person}{Yue Liu}, \bibinfo{person}{Wenxuan Tu}, \bibinfo{person}{Siwei Wang}, \bibinfo{person}{Sihang Zhou}, \bibinfo{person}{Xinwang Liu}, {and} \bibinfo{person}{Fuchun Sun}.} \bibinfo{year}{2022}\natexlab{}.
\newblock \showarticletitle{Reasoning over different types of knowledge graphs: Static, temporal and multi-modal}.
\newblock \bibinfo{journal}{\emph{arXiv preprint arXiv:2212.05767}} (\bibinfo{year}{2022}).
\newblock


\bibitem[Liu et~al\mbox{.}(2019)]%
        {DBLP:conf/esws/LiuLGNOR19}
\bibfield{author}{\bibinfo{person}{Ye Liu}, \bibinfo{person}{Hui Li}, \bibinfo{person}{Alberto Garc{\'{\i}}a{-}Dur{\'{a}}n}, \bibinfo{person}{Mathias Niepert}, \bibinfo{person}{Daniel O{\~{n}}oro{-}Rubio}, {and} \bibinfo{person}{David~S. Rosenblum}.} \bibinfo{year}{2019}\natexlab{}.
\newblock \showarticletitle{{MMKG:} Multi-modal Knowledge Graphs}. In \bibinfo{booktitle}{\emph{The Semantic Web - 16th International Conference, {ESWC} 2019, Portoro{\v{z}}, Slovenia, June 2-6, 2019, Proceedings}} \emph{(\bibinfo{series}{Lecture Notes in Computer Science}, Vol.~\bibinfo{volume}{11503})}, \bibfield{editor}{\bibinfo{person}{Pascal Hitzler}, \bibinfo{person}{Miriam Fern{\'{a}}ndez}, \bibinfo{person}{Krzysztof Janowicz}, \bibinfo{person}{Amrapali Zaveri}, \bibinfo{person}{Alasdair J.~G. Gray}, \bibinfo{person}{Vanessa L{\'{o}}pez}, \bibinfo{person}{Armin Haller}, {and} \bibinfo{person}{Karl Hammar}} (Eds.). \bibinfo{publisher}{Springer}, \bibinfo{pages}{459--474}.
\newblock
\urldef\tempurl%
\url{https://doi.org/10.1007/978-3-030-21348-0\_30}
\showDOI{\tempurl}


\bibitem[Loshchilov and Hutter(2019)]%
        {DBLP:conf/iclr/LoshchilovH19}
\bibfield{author}{\bibinfo{person}{Ilya Loshchilov} {and} \bibinfo{person}{Frank Hutter}.} \bibinfo{year}{2019}\natexlab{}.
\newblock \showarticletitle{Decoupled Weight Decay Regularization}. In \bibinfo{booktitle}{\emph{7th International Conference on Learning Representations, {ICLR} 2019, New Orleans, LA, USA, May 6-9, 2019}}. \bibinfo{publisher}{OpenReview.net}.
\newblock
\urldef\tempurl%
\url{https://openreview.net/forum?id=Bkg6RiCqY7}
\showURL{%
\tempurl}


\bibitem[Lu et~al\mbox{.}(2019)]%
        {DBLP:conf/nips/LuBPL19}
\bibfield{author}{\bibinfo{person}{Jiasen Lu}, \bibinfo{person}{Dhruv Batra}, \bibinfo{person}{Devi Parikh}, {and} \bibinfo{person}{Stefan Lee}.} \bibinfo{year}{2019}\natexlab{}.
\newblock \showarticletitle{ViLBERT: Pretraining Task-Agnostic Visiolinguistic Representations for Vision-and-Language Tasks}. In \bibinfo{booktitle}{\emph{Advances in Neural Information Processing Systems 32: Annual Conference on Neural Information Processing Systems 2019, NeurIPS 2019, December 8-14, 2019, Vancouver, BC, Canada}}, \bibfield{editor}{\bibinfo{person}{Hanna~M. Wallach}, \bibinfo{person}{Hugo Larochelle}, \bibinfo{person}{Alina Beygelzimer}, \bibinfo{person}{Florence d'Alch{\'{e}}{-}Buc}, \bibinfo{person}{Emily~B. Fox}, {and} \bibinfo{person}{Roman Garnett}} (Eds.). \bibinfo{pages}{13--23}.
\newblock
\urldef\tempurl%
\url{https://proceedings.neurips.cc/paper/2019/hash/c74d97b01eae257e44aa9d5bade97baf-Abstract.html}
\showURL{%
\tempurl}


\bibitem[Lully et~al\mbox{.}(2018)]%
        {DBLP:conf/i-semantics/LullyLSR18}
\bibfield{author}{\bibinfo{person}{Vincent Lully}, \bibinfo{person}{Philippe Laublet}, \bibinfo{person}{Milan Stankovic}, {and} \bibinfo{person}{Filip Radulovic}.} \bibinfo{year}{2018}\natexlab{}.
\newblock \showarticletitle{Exploring the synergy between knowledge graph and computer vision for personalisation systems}. In \bibinfo{booktitle}{\emph{Proceedings of the 14th International Conference on Semantic Systems, SEMANTiCS 2018, Vienna, Austria, September 10-13, 2018}} \emph{(\bibinfo{series}{Procedia Computer Science}, Vol.~\bibinfo{volume}{137})}, \bibfield{editor}{\bibinfo{person}{Anna Fensel}, \bibinfo{person}{Victor de~Boer}, \bibinfo{person}{Tassilo Pellegrini}, \bibinfo{person}{Elmar Kiesling}, \bibinfo{person}{Bernhard Haslhofer}, \bibinfo{person}{Laura Hollink}, {and} \bibinfo{person}{Alexander Schindler}} (Eds.). \bibinfo{publisher}{Elsevier}, \bibinfo{pages}{175--186}.
\newblock
\urldef\tempurl%
\url{https://doi.org/10.1016/j.procs.2018.09.017}
\showDOI{\tempurl}


\bibitem[Mitra et~al\mbox{.}(2022)]%
        {DBLP:conf/naacl/MitraRS22}
\bibfield{author}{\bibinfo{person}{Sayantan Mitra}, \bibinfo{person}{Roshni~R. Ramnani}, {and} \bibinfo{person}{Shubhashis Sengupta}.} \bibinfo{year}{2022}\natexlab{}.
\newblock \showarticletitle{Constraint-based Multi-hop Question Answering with Knowledge Graph}. In \bibinfo{booktitle}{\emph{Proceedings of the 2022 Conference of the North American Chapter of the Association for Computational Linguistics: Human Language Technologies: Industry Track, {NAACL} 2022, Hybrid: Seattle, Washington, {USA} + Online, July 10-15, 2022}}, \bibfield{editor}{\bibinfo{person}{Anastassia Loukina}, \bibinfo{person}{Rashmi Gangadharaiah}, {and} \bibinfo{person}{Bonan Min}} (Eds.). \bibinfo{publisher}{Association for Computational Linguistics}, \bibinfo{pages}{280--288}.
\newblock
\urldef\tempurl%
\url{https://doi.org/10.18653/v1/2022.naacl-industry.31}
\showDOI{\tempurl}


\bibitem[Mokady et~al\mbox{.}(2021)]%
        {DBLP:journals/corr/abs-2111-09734}
\bibfield{author}{\bibinfo{person}{Ron Mokady}, \bibinfo{person}{Amir Hertz}, {and} \bibinfo{person}{Amit~H. Bermano}.} \bibinfo{year}{2021}\natexlab{}.
\newblock \showarticletitle{ClipCap: {CLIP} Prefix for Image Captioning}.
\newblock \bibinfo{journal}{\emph{CoRR}}  \bibinfo{volume}{abs/2111.09734} (\bibinfo{year}{2021}).
\newblock
\showeprint[arXiv]{2111.09734}
\urldef\tempurl%
\url{https://arxiv.org/abs/2111.09734}
\showURL{%
\tempurl}


\bibitem[Shen et~al\mbox{.}(2022)]%
        {DBLP:conf/iclr/ShenLTBRCYK22}
\bibfield{author}{\bibinfo{person}{Sheng Shen}, \bibinfo{person}{Liunian~Harold Li}, \bibinfo{person}{Hao Tan}, \bibinfo{person}{Mohit Bansal}, \bibinfo{person}{Anna Rohrbach}, \bibinfo{person}{Kai{-}Wei Chang}, \bibinfo{person}{Zhewei Yao}, {and} \bibinfo{person}{Kurt Keutzer}.} \bibinfo{year}{2022}\natexlab{}.
\newblock \showarticletitle{How Much Can {CLIP} Benefit Vision-and-Language Tasks?}. In \bibinfo{booktitle}{\emph{The Tenth International Conference on Learning Representations, {ICLR} 2022, Virtual Event, April 25-29, 2022}}. \bibinfo{publisher}{OpenReview.net}.
\newblock
\urldef\tempurl%
\url{https://openreview.net/forum?id=zf\_Ll3HZWgy}
\showURL{%
\tempurl}


\bibitem[Tang et~al\mbox{.}(2020)]%
        {DBLP:conf/cvpr/TangNHSZ20}
\bibfield{author}{\bibinfo{person}{Kaihua Tang}, \bibinfo{person}{Yulei Niu}, \bibinfo{person}{Jianqiang Huang}, \bibinfo{person}{Jiaxin Shi}, {and} \bibinfo{person}{Hanwang Zhang}.} \bibinfo{year}{2020}\natexlab{}.
\newblock \showarticletitle{Unbiased Scene Graph Generation From Biased Training}. In \bibinfo{booktitle}{\emph{2020 {IEEE/CVF} Conference on Computer Vision and Pattern Recognition, {CVPR} 2020, Seattle, WA, USA, June 13-19, 2020}}. \bibinfo{publisher}{Computer Vision Foundation / {IEEE}}, \bibinfo{pages}{3713--3722}.
\newblock
\urldef\tempurl%
\url{https://doi.org/10.1109/CVPR42600.2020.00377}
\showDOI{\tempurl}


\bibitem[Vaswani et~al\mbox{.}(2017)]%
        {DBLP:conf/nips/VaswaniSPUJGKP17}
\bibfield{author}{\bibinfo{person}{Ashish Vaswani}, \bibinfo{person}{Noam Shazeer}, \bibinfo{person}{Niki Parmar}, \bibinfo{person}{Jakob Uszkoreit}, \bibinfo{person}{Llion Jones}, \bibinfo{person}{Aidan~N. Gomez}, \bibinfo{person}{Lukasz Kaiser}, {and} \bibinfo{person}{Illia Polosukhin}.} \bibinfo{year}{2017}\natexlab{}.
\newblock \showarticletitle{Attention is All you Need}. In \bibinfo{booktitle}{\emph{Advances in Neural Information Processing Systems 30: Annual Conference on Neural Information Processing Systems 2017, December 4-9, 2017, Long Beach, CA, {USA}}}, \bibfield{editor}{\bibinfo{person}{Isabelle Guyon}, \bibinfo{person}{Ulrike von Luxburg}, \bibinfo{person}{Samy Bengio}, \bibinfo{person}{Hanna~M. Wallach}, \bibinfo{person}{Rob Fergus}, \bibinfo{person}{S.~V.~N. Vishwanathan}, {and} \bibinfo{person}{Roman Garnett}} (Eds.). \bibinfo{pages}{5998--6008}.
\newblock
\urldef\tempurl%
\url{https://proceedings.neurips.cc/paper/2017/hash/3f5ee243547dee91fbd053c1c4a845aa-Abstract.html}
\showURL{%
\tempurl}


\bibitem[Wan et~al\mbox{.}(2021)]%
        {DBLP:conf/aaai/WanZDHYP21}
\bibfield{author}{\bibinfo{person}{Hai Wan}, \bibinfo{person}{Manrong Zhang}, \bibinfo{person}{Jianfeng Du}, \bibinfo{person}{Ziling Huang}, \bibinfo{person}{Yufei Yang}, {and} \bibinfo{person}{Jeff~Z. Pan}.} \bibinfo{year}{2021}\natexlab{}.
\newblock \showarticletitle{{FL-MSRE:} {A} Few-Shot Learning based Approach to Multimodal Social Relation Extraction}. In \bibinfo{booktitle}{\emph{Thirty-Fifth {AAAI} Conference on Artificial Intelligence, {AAAI} 2021, Thirty-Third Conference on Innovative Applications of Artificial Intelligence, {IAAI} 2021, The Eleventh Symposium on Educational Advances in Artificial Intelligence, {EAAI} 2021, Virtual Event, February 2-9, 2021}}. \bibinfo{publisher}{{AAAI} Press}, \bibinfo{pages}{13916--13923}.
\newblock
\urldef\tempurl%
\url{https://ojs.aaai.org/index.php/AAAI/article/view/17639}
\showURL{%
\tempurl}


\bibitem[Wan et~al\mbox{.}(2022)]%
        {DBLP:conf/mm/WanLLLWZ22}
\bibfield{author}{\bibinfo{person}{Xinhang Wan}, \bibinfo{person}{Jiyuan Liu}, \bibinfo{person}{Weixuan Liang}, \bibinfo{person}{Xinwang Liu}, \bibinfo{person}{Yi Wen}, {and} \bibinfo{person}{En Zhu}.} \bibinfo{year}{2022}\natexlab{}.
\newblock \showarticletitle{Continual Multi-view Clustering}. In \bibinfo{booktitle}{\emph{{MM} '22: The 30th {ACM} International Conference on Multimedia, Lisboa, Portugal, October 10 - 14, 2022}}, \bibfield{editor}{\bibinfo{person}{Jo{\~{a}}o Magalh{\~{a}}es}, \bibinfo{person}{Alberto~Del Bimbo}, \bibinfo{person}{Shin'ichi Satoh}, \bibinfo{person}{Nicu Sebe}, \bibinfo{person}{Xavier Alameda{-}Pineda}, \bibinfo{person}{Qin Jin}, \bibinfo{person}{Vincent Oria}, {and} \bibinfo{person}{Laura Toni}} (Eds.). \bibinfo{publisher}{{ACM}}, \bibinfo{pages}{3676--3684}.
\newblock
\urldef\tempurl%
\url{https://doi.org/10.1145/3503161.3547864}
\showDOI{\tempurl}


\bibitem[Wan et~al\mbox{.}(2023)]%
        {wan2023auto}
\bibfield{author}{\bibinfo{person}{Xinhang Wan}, \bibinfo{person}{Xinwang Liu}, \bibinfo{person}{Jiyuan Liu}, \bibinfo{person}{Siwei Wang}, \bibinfo{person}{Yi Wen}, \bibinfo{person}{Weixuan Liang}, \bibinfo{person}{En Zhu}, \bibinfo{person}{Zhe Liu}, {and} \bibinfo{person}{Lu Zhou}.} \bibinfo{year}{2023}\natexlab{}.
\newblock \showarticletitle{Auto-weighted multi-view clustering for large-scale data}.
\newblock \bibinfo{journal}{\emph{arXiv preprint arXiv:2303.01983}} (\bibinfo{year}{2023}).
\newblock


\bibitem[Wang et~al\mbox{.}(2020)]%
        {DBLP:journals/bdr/WangWQZ20}
\bibfield{author}{\bibinfo{person}{Meng Wang}, \bibinfo{person}{Haofen Wang}, \bibinfo{person}{Guilin Qi}, {and} \bibinfo{person}{Qiushuo Zheng}.} \bibinfo{year}{2020}\natexlab{}.
\newblock \showarticletitle{Richpedia: {A} Large-Scale, Comprehensive Multi-Modal Knowledge Graph}.
\newblock \bibinfo{journal}{\emph{Big Data Res.}}  \bibinfo{volume}{22} (\bibinfo{year}{2020}), \bibinfo{pages}{100159}.
\newblock
\urldef\tempurl%
\url{https://doi.org/10.1016/j.bdr.2020.100159}
\showDOI{\tempurl}


\bibitem[Wang et~al\mbox{.}(2022a)]%
        {DBLP:conf/emnlp/WangCJXTL22}
\bibfield{author}{\bibinfo{person}{Xinyu Wang}, \bibinfo{person}{Jiong Cai}, \bibinfo{person}{Yong Jiang}, \bibinfo{person}{Pengjun Xie}, \bibinfo{person}{Kewei Tu}, {and} \bibinfo{person}{Wei Lu}.} \bibinfo{year}{2022}\natexlab{a}.
\newblock \showarticletitle{Named Entity and Relation Extraction with Multi-Modal Retrieval}. In \bibinfo{booktitle}{\emph{Findings of the Association for Computational Linguistics: {EMNLP} 2022, Abu Dhabi, United Arab Emirates, December 7-11, 2022}}, \bibfield{editor}{\bibinfo{person}{Yoav Goldberg}, \bibinfo{person}{Zornitsa Kozareva}, {and} \bibinfo{person}{Yue Zhang}} (Eds.). \bibinfo{publisher}{Association for Computational Linguistics}, \bibinfo{pages}{5925--5936}.
\newblock
\urldef\tempurl%
\url{https://aclanthology.org/2022.findings-emnlp.437}
\showURL{%
\tempurl}


\bibitem[Wang et~al\mbox{.}(2022b)]%
        {DBLP:conf/icmcs/WangYLTJYZX22}
\bibfield{author}{\bibinfo{person}{Xuwu Wang}, \bibinfo{person}{Jiabo Ye}, \bibinfo{person}{Zhixu Li}, \bibinfo{person}{Junfeng Tian}, \bibinfo{person}{Yong Jiang}, \bibinfo{person}{Ming Yan}, \bibinfo{person}{Ji Zhang}, {and} \bibinfo{person}{Yanghua Xiao}.} \bibinfo{year}{2022}\natexlab{b}.
\newblock \showarticletitle{{CAT-MNER:} Multimodal Named Entity Recognition with Knowledge-Refined Cross-Modal Attention}. In \bibinfo{booktitle}{\emph{{IEEE} International Conference on Multimedia and Expo, {ICME} 2022, Taipei, Taiwan, July 18-22, 2022}}. \bibinfo{publisher}{{IEEE}}, \bibinfo{pages}{1--6}.
\newblock
\urldef\tempurl%
\url{https://doi.org/10.1109/ICME52920.2022.9859972}
\showDOI{\tempurl}


\bibitem[Wu et~al\mbox{.}(2022)]%
        {wu2022state}
\bibfield{author}{\bibinfo{person}{Yuxia Wu}, \bibinfo{person}{Lizi Liao}, \bibinfo{person}{Gangyi Zhang}, \bibinfo{person}{Wenqiang Lei}, \bibinfo{person}{Guoshuai Zhao}, \bibinfo{person}{Xueming Qian}, {and} \bibinfo{person}{Tat-Seng Chua}.} \bibinfo{year}{2022}\natexlab{}.
\newblock \showarticletitle{State graph reasoning for multimodal conversational recommendation}.
\newblock \bibinfo{journal}{\emph{IEEE Transactions on Multimedia}} (\bibinfo{year}{2022}).
\newblock


\bibitem[Xu et~al\mbox{.}(2022)]%
        {DBLP:conf/coling/0023HDWSSX22}
\bibfield{author}{\bibinfo{person}{Bo Xu}, \bibinfo{person}{Shizhou Huang}, \bibinfo{person}{Ming Du}, \bibinfo{person}{Hongya Wang}, \bibinfo{person}{Hui Song}, \bibinfo{person}{Chaofeng Sha}, {and} \bibinfo{person}{Yanghua Xiao}.} \bibinfo{year}{2022}\natexlab{}.
\newblock \showarticletitle{Different Data, Different Modalities! Reinforced Data Splitting for Effective Multimodal Information Extraction from Social Media Posts}. In \bibinfo{booktitle}{\emph{Proceedings of the 29th International Conference on Computational Linguistics, {COLING} 2022, Gyeongju, Republic of Korea, October 12-17, 2022}}, \bibfield{editor}{\bibinfo{person}{Nicoletta Calzolari}, \bibinfo{person}{Chu{-}Ren Huang}, \bibinfo{person}{Hansaem Kim}, \bibinfo{person}{James Pustejovsky}, \bibinfo{person}{Leo Wanner}, \bibinfo{person}{Key{-}Sun Choi}, \bibinfo{person}{Pum{-}Mo Ryu}, \bibinfo{person}{Hsin{-}Hsi Chen}, \bibinfo{person}{Lucia Donatelli}, \bibinfo{person}{Heng Ji}, \bibinfo{person}{Sadao Kurohashi}, \bibinfo{person}{Patrizia Paggio}, \bibinfo{person}{Nianwen Xue}, \bibinfo{person}{Seokhwan Kim}, \bibinfo{person}{Younggyun Hahm}, \bibinfo{person}{Zhong He}, \bibinfo{person}{Tony~Kyungil Lee}, \bibinfo{person}{Enrico Santus}, \bibinfo{person}{Francis Bond}, {and}
  \bibinfo{person}{Seung{-}Hoon Na}} (Eds.). \bibinfo{publisher}{International Committee on Computational Linguistics}, \bibinfo{pages}{1855--1864}.
\newblock
\urldef\tempurl%
\url{https://aclanthology.org/2022.coling-1.160}
\showURL{%
\tempurl}


\bibitem[Zeng et~al\mbox{.}(2022)]%
        {DBLP:conf/icml/ZengZL22}
\bibfield{author}{\bibinfo{person}{Yan Zeng}, \bibinfo{person}{Xinsong Zhang}, {and} \bibinfo{person}{Hang Li}.} \bibinfo{year}{2022}\natexlab{}.
\newblock \showarticletitle{Multi-Grained Vision Language Pre-Training: Aligning Texts with Visual Concepts}. In \bibinfo{booktitle}{\emph{International Conference on Machine Learning, {ICML} 2022, 17-23 July 2022, Baltimore, Maryland, {USA}}} \emph{(\bibinfo{series}{Proceedings of Machine Learning Research}, Vol.~\bibinfo{volume}{162})}, \bibfield{editor}{\bibinfo{person}{Kamalika Chaudhuri}, \bibinfo{person}{Stefanie Jegelka}, \bibinfo{person}{Le~Song}, \bibinfo{person}{Csaba Szepesv{\'{a}}ri}, \bibinfo{person}{Gang Niu}, {and} \bibinfo{person}{Sivan Sabato}} (Eds.). \bibinfo{publisher}{{PMLR}}, \bibinfo{pages}{25994--26009}.
\newblock
\urldef\tempurl%
\url{https://proceedings.mlr.press/v162/zeng22c.html}
\showURL{%
\tempurl}


\bibitem[Zhang et~al\mbox{.}(2017)]%
        {DBLP:conf/emnlp/ZhangZCAM17}
\bibfield{author}{\bibinfo{person}{Yuhao Zhang}, \bibinfo{person}{Victor Zhong}, \bibinfo{person}{Danqi Chen}, \bibinfo{person}{Gabor Angeli}, {and} \bibinfo{person}{Christopher~D. Manning}.} \bibinfo{year}{2017}\natexlab{}.
\newblock \showarticletitle{Position-aware Attention and Supervised Data Improve Slot Filling}. In \bibinfo{booktitle}{\emph{Proceedings of the 2017 Conference on Empirical Methods in Natural Language Processing, {EMNLP} 2017, Copenhagen, Denmark, September 9-11, 2017}}, \bibfield{editor}{\bibinfo{person}{Martha Palmer}, \bibinfo{person}{Rebecca Hwa}, {and} \bibinfo{person}{Sebastian Riedel}} (Eds.). \bibinfo{publisher}{Association for Computational Linguistics}, \bibinfo{pages}{35--45}.
\newblock
\urldef\tempurl%
\url{https://doi.org/10.18653/v1/d17-1004}
\showDOI{\tempurl}


\bibitem[Zhao et~al\mbox{.}(2023)]%
        {zhao2023tsvfn}
\bibfield{author}{\bibinfo{person}{Qihui Zhao}, \bibinfo{person}{Tianhan Gao}, {and} \bibinfo{person}{Nan Guo}.} \bibinfo{year}{2023}\natexlab{}.
\newblock \showarticletitle{TSVFN: Two-Stage Visual Fusion Network for multimodal relation extraction}.
\newblock \bibinfo{journal}{\emph{Information Processing \& Management}} \bibinfo{volume}{60}, \bibinfo{number}{3} (\bibinfo{year}{2023}), \bibinfo{pages}{103264}.
\newblock


\bibitem[Zheng et~al\mbox{.}(2021a)]%
        {DBLP:conf/mm/ZhengFFCL021}
\bibfield{author}{\bibinfo{person}{Changmeng Zheng}, \bibinfo{person}{Junhao Feng}, \bibinfo{person}{Ze Fu}, \bibinfo{person}{Yi Cai}, \bibinfo{person}{Qing Li}, {and} \bibinfo{person}{Tao Wang}.} \bibinfo{year}{2021}\natexlab{a}.
\newblock \showarticletitle{Multimodal Relation Extraction with Efficient Graph Alignment}. In \bibinfo{booktitle}{\emph{{MM} '21: {ACM} Multimedia Conference, Virtual Event, China, October 20 - 24, 2021}}, \bibfield{editor}{\bibinfo{person}{Heng~Tao Shen}, \bibinfo{person}{Yueting Zhuang}, \bibinfo{person}{John~R. Smith}, \bibinfo{person}{Yang Yang}, \bibinfo{person}{Pablo C{\'{e}}sar}, \bibinfo{person}{Florian Metze}, {and} \bibinfo{person}{Balakrishnan Prabhakaran}} (Eds.). \bibinfo{publisher}{{ACM}}, \bibinfo{pages}{5298--5306}.
\newblock
\urldef\tempurl%
\url{https://doi.org/10.1145/3474085.3476968}
\showDOI{\tempurl}


\bibitem[Zheng et~al\mbox{.}(2021b)]%
        {DBLP:conf/icmcs/ZhengWFF021}
\bibfield{author}{\bibinfo{person}{Changmeng Zheng}, \bibinfo{person}{Zhiwei Wu}, \bibinfo{person}{Junhao Feng}, \bibinfo{person}{Ze Fu}, {and} \bibinfo{person}{Yi Cai}.} \bibinfo{year}{2021}\natexlab{b}.
\newblock \showarticletitle{{MNRE:} {A} Challenge Multimodal Dataset for Neural Relation Extraction with Visual Evidence in Social Media Posts}. In \bibinfo{booktitle}{\emph{2021 {IEEE} International Conference on Multimedia and Expo, {ICME} 2021, Shenzhen, China, July 5-9, 2021}}. \bibinfo{publisher}{{IEEE}}, \bibinfo{pages}{1--6}.
\newblock
\urldef\tempurl%
\url{https://doi.org/10.1109/ICME51207.2021.9428274}
\showDOI{\tempurl}


\bibitem[Zheng et~al\mbox{.}(2023)]%
        {DBLP:conf/icde/Zheng0QYCZ23}
\bibfield{author}{\bibinfo{person}{Shangfei Zheng}, \bibinfo{person}{Weiqing Wang}, \bibinfo{person}{Jianfeng Qu}, \bibinfo{person}{Hongzhi Yin}, \bibinfo{person}{Wei Chen}, {and} \bibinfo{person}{Lei Zhao}.} \bibinfo{year}{2023}\natexlab{}.
\newblock \showarticletitle{{MMKGR:} Multi-hop Multi-modal Knowledge Graph Reasoning}. In \bibinfo{booktitle}{\emph{39th {IEEE} International Conference on Data Engineering, {ICDE} 2023, Anaheim, CA, USA, April 3-7, 2023}}. \bibinfo{publisher}{{IEEE}}, \bibinfo{pages}{96--109}.
\newblock
\urldef\tempurl%
\url{https://doi.org/10.1109/ICDE55515.2023.00015}
\showDOI{\tempurl}


\end{thebibliography}

\appendix

\section{MORE Dataset Construction Details}
\label{dataset_construction}

The complete procedure can be partitioned into two steps: 1) generating a comprehensive pool of candidate textual news titles with corresponding images from The New York Times and Yahoo News, followed by a meticulous cleansing process, and 2) devising an annotation mechanism that extracts the entities from the news titles as well as objects from the images, followed by identifying their relations. This involves leveraging human annotators and heuristic methods to achieve an accurate and refined multimodal object-entity relation extraction dataset.

\subsection{Data Collection}

For our dataset construction, we obtained data primarily from The New York Times English news and Yahoo News ranging from 2019 to 2022. Each data instance contains a textual news title and a corresponding news image, and covers a wide range of topics in fields such as sports and politics. After filtering out unqualified data, including irrelevant texts and images, textual news titles that do not contain any entities, and images that do not contain any objects, we obtained a candidate set of 15,000 multimodal news data. This set of filtered data has been meticulously selected to ensure its relevance and appropriateness for our research purposes.

\subsection{Data Annotation}

The candidate multimodal news we collected was annotated in three distinct stages. Firstly, entities in the textual news titles and objects in images were automatically identified, and the results were manually reviewed and corrected. Secondly, relations between entities and objects were manually annotated. Finally, overlapping objects mentioned in the text were filtered out.

\textbf{Stage 1: Entity Identification and Object Detection.} We utilized the AllenNLP named entity recognition tool to identify and label the entities in textual news titles. Moreover, we employed the Yolo V5 object detection tool to detect object areas in the corresponding news images. 
Considering the possibility of omissions and errors by both tools, all extracted objects and entities were reviewed and corrected manually by our annotators. To enhance the efficiency of the annotation process, we designed and developed an annotation tool that presented the entities in the text and the corresponding objects' area in the image for the annotators to reference. Overall, this approach ensured that our annotations were of high quality and accurately reflected the entities and objects present in the dataset.

\textbf{Stage 2: Object-Entity Relation Annotation.} We recruited well-educated annotators to manually annotate the relations between the entities and objects. Each sample of our dataset was annotated by at least two annotators. To ensure an unbiased annotation process, relations were randomly assigned to annotators from the candidate set. Each annotator was responsible for examining the textual titles and images and deducing the relations between the entities and objects. The data did not clearly indicate any pre-defined relations will be labeled as \verb|none|. To enhance the quality of our annotations, we used the \verb|cohen_kappa_score| function from the sklearn library to calculate the inter-annotator agreement, which resulted in a Kappa value of 0.7185. In cases where there were discrepancies or conflicts in the annotations, a third annotator was consulted, and they will discuss to reach an agreement. 

\textbf{Stage 3: Object-Overlapped Data Filtering.} In order to refine the scope of multimodal object-entity relation extraction task, we limited our focus to relations in which visual objects did not co-occur with any entities mentioned in the textual news titles. To accomplish this, we utilized a visual entity linking tool to identify the names of visual objects and subsequently removed any objects that appeared in the textual news titles. Following this filtering process, we were left with a high-quality dataset of over 3,000 news articles and more than 20,000 object-entity relational facts, out of the original 15,000 multimodal news data.
This approach ensured that our dataset focused on relatable object-entity relationships illustrated in images, rather than those that were already mentioned explicitly in the textual news titles, resulting in a more relevant set of data for our multimodal object-entity relation extraction task.

\subsection{Annotators}

We hired a team of professional annotators, comprising two dedicated to the annotation process and one consultant, who worked diligently for three months. Each individual sample annotation was priced at 2 RMB, which is approximately equivalent to 0.28 USD. To ensure that the annotators were well-trained, we adopted a principled training procedure and required them to pass test tasks before proceeding to the actual annotation of the dataset. This approach ensured our annotators had a solid understanding of the task at hand and produced high-quality annotations.

\end{document}